\documentclass[12pt,a4paper]{article}

\usepackage[left=2.5cm,right=2.5cm,top=3cm,bottom=3cm,bindingoffset=0cm]{geometry}

\usepackage{amsmath}
\usepackage{amsfonts}
\usepackage{amssymb}
\usepackage{graphicx}
\usepackage{url}
\usepackage{natbib}		
\bibpunct{(}{)}{;}{a}{,}{,}
\usepackage[small,bf]{caption}
\usepackage{framed}
\usepackage{multirow}
\usepackage{arydshln}
\usepackage{soul}

\usepackage{ulem}
\usepackage{color}

\usepackage{amsthm}
\theoremstyle{definition}
\newtheorem{remark}{Remark}
\theoremstyle{definition}
\newtheorem{example}{Example}

\newcommand{\Ch}{ C^{\text{He}} }

\newcommand{\Var}{\mathrm{Var}}
\newcommand{\R}{\mathbb{R}}
\newcommand{\E}{\mathbb{E}}

\newcommand{\Pconv}{ \overset{P}{\rightarrow} }
\DeclareMathOperator*{\argmax}{arg\,max}
\DeclareMathOperator*{\essinf}{ess\,inf}
\DeclareMathOperator*{\esssup}{ess\,sup}

\newcommand{\ru}{r^u(u_t)}
\newcommand{\ku}{\kappa^u(u_t)}
\newcommand{\tu}{\theta^u(u_t)}

\makeatletter
\renewcommand*{\@fnsymbol}[1]{\ensuremath{\ifcase#1\or 1\or 2\or \ast\or    \dagger\or  \ddagger\or 
    \mathsection\or \mathparagraph\or \|\or **\or \dagger\dagger
    \or \ddagger\ddagger \else\@ctrerr\fi}}
\makeatother

\title{\sc European Option Pricing with Stochastic Volatility models under Parameter Uncertainty}
\author{Samuel N. Cohen\thanks{Mathematical Institute, University of Oxford.}\, and Martin Tegn\'{e}r$^{1,}$\thanks{Department of Engineering Science, University of Oxford and Department of Mathematical Sciences, University of Copenhagen.} $^,$\thanks{Corresponding author.} \\
{\small  samuel.cohen@maths.ox.ac.uk,  martin.tegner@eng.ox.ac.uk}
}
\date{}

\begin{document}

\maketitle
\begin{abstract}
\noindent We consider stochastic volatility models under parameter uncertainty and investigate how model derived prices of European options are affected. We let the pricing parameters evolve dynamically in time within a specified region, and formalise the problem as a control problem where the control acts on the parameters to maximise/minimise the option value. Through a dual representation with backward stochastic differential equations, we obtain explicit equations for Heston's model and investigate several numerical solutions thereof. In an empirical study, we apply our results to market data from the S\&P 500 index where the model is estimated to historical asset prices. We find that the conservative model-prices cover 98\% of  the considered market-prices for a set of European call options. 
\end{abstract}

\section{Introduction}

In this paper, we consider the problem of European-option pricing when the underlying assets are assumed to follow a stochastic volatility model in a setting that accommodates for  
\textit{parameter uncertainty}, and in particular, how this transfers to \textit{conservative bounds} for derived option prices.

Stochastic volatility models feature an instantaneous variance of the asset price, the \textit{volatility}, that evolves stochastically in time. It is a natural generalisation of the seminal constant-volatility model of \cite{black1973pricing}, and examples include the models introduced by \cite{hull1987pricing}, \cite{stein1991stock}, \cite{heston1993closed}, \cite{bates1996jumps} and \cite{heston1997simple} to mention a few. Evidence supporting this generalisation in terms of empirical asset-return behaviour goes back to \cite{black1976stuedies}, while for instance \cite{stein1989overreactions} highlights the prediction mismatch of a constant-volatility models and option prices observed from the market. Stochastic volatility serves as an attractive alternative and numerous studies are available from the literature in their favour.

Being a parametric model immediately implies that the stochastic volatility model has to be fitted with data before it is employed for pricing or hedging market instruments. At least two approaches are conventional for this purpose: either estimation from historical asset-prices, or calibration from market option-prices by matching the model derived price (or a combination of the two, see for example \cite{ai2007maximum}). 
Regardless of which approach is used, one is exposed to \textit{parameter uncertainty} since point-estimates from either are subject to errors. An estimator based on observed time-series of asset prices has an inherent variance (and could potentially be biased), while the calibration problem might be ill-posed---the minimum obtained from numerical optimisation can be local, and several parameter settings might give the same model-to-market matching.

A concept of parameter uncertainty naturally arises with statistical estimation from asset-prices since the error of the point-estimate can be quantified by an inferred confidence interval. The confidence interval thus defines an uncertainty set which contains the true value of the model parameters, at a given confidence level. In this case, inferred uncertainty and 
estimated parameters will be associated with the \textit{real-world} probability measure, as opposed to the \textit{risk-neutral} measure(s) used for no-arbitrage pricing. 
On the other hand, calibration from option-prices will give a set of parameters associated with the risk-neutral measure. In this case, however, there is no obvious way of how to deduce an uncertainty set for the parameters which quantifies the errors that stem from the calibration.

{The question remains how the parameter uncertainty affects option prices as outputted by the stochastic volatility model.  In the case of statistical estimation, one needs to establish the relation between the parameters under the statistical measure and under a risk-neutral pricing measure. In financial terms, this is accommodated by the market price of risk, and typically in such a way that the model remains form-invariant. The uncertainty may then be propagated to the risk-neutral parameters which are used for option pricing. We consider uncertainty in drift- and jump parameters}\footnote{A supporting case for this assumption is the fact pointed out for instance by \cite{rogers2001relaxed}: while volatilities may be estimated within reasonable confidence with a few years of data, drift estimation requires data from much longer time periods.} {to  offer an interpretation of the parameter uncertainty as representative for the  \textit{incompleteness} of the stochastic volatility model: there exists a space of equivalent pricing measures as given by the span of risk-neutral parameter values in the uncertainty set (we elaborate on this in the introducing discussion of Section} \ref{PIsecE}).

We immediately look at the model pricing from a best/worst case point of view, and aim to obtain conservative \textit{pricing bounds} inferred from the parameter uncertainty. Two approaches are fair: either optimising the pricing function over the parameters constrained by the uncertainty set, or treating the parameters as dynamical components of a \textit{control process} which acts to optimise the option value. The former is thus a special case of the latter where the control process is restricted to take constant values only. We formalise the problem as a control problem and since all pricing measures are equivalent, this can be seen as change of measure problem. 
Following the results due to \cite{quenez1997stochastic}, the optimal value function of the option price may then by expressed as a backward stochastic differential equation.

The postulation of parameter uncertainty, or more generally \textit{model uncertainty}, as an inherent model feature is certainly not novel and its importance in finance was early acknowledged 
by \cite{dermanMR}. Conceptually, model uncertainty draws on the principles due to \cite{keynes1921treatise} and \cite{knight1921risk} of the \textit{unknown unknown} as distinguished from the \textit{known unknown}. Following the overview by \cite{bannor2014model}, model uncertainty---the unknown unknown---refers to the situation when a whole family of models is available for the financial market, but the \textit{likelihood} of each individual model is unknown. Parameter uncertainty is thus the special case where the family of models may be parametrized. Further, if a probability measure is 
attributed to the model (parameter) family---the known unknown---one is in the situation of \textit{model (parameter) risk}.

When it comes to option pricing, Bayesian methods offer a fruitful way of inferring parameter and model risk, and take it into account by model averaging,
see for example \cite{jacquier2000bayesian}, \cite{bunnin2002option}, \cite{gupta2010model} and the non-parametric approach to local volatility by \cite{mtGPLV}.
 Considering the situation of model uncertainty, the worst-case approach taken here was pioneered in the works of \cite{el1995dynamic}, \cite{avellaneda1995pricing}, \cite{lyons1995uncertain} and \cite{avellaneda1996managing}. Our control-theoretic approach is similar to that of \citeauthor{avellaneda1995pricing} but in contrast to their unspecified volatility, we place ourself in a ``within-model'' setting of parametrised volatility models where the parameters are controlled instead of the volatility itself. 
We thus account for a case 
where the uncertain family of volatility models gives a more detailed description of the financial market. Arguably, this implies conservative prices which are more realistic. Since we also suggest how to infer the uncertainty set for the parameters, our approach should be particularly appealing for stochastic-volatility inclined practitioners.

\vspace{10pt}

\noindent\textbf{Overview.} The model proposed by \cite{heston1993closed} will be the working model of our study, and we present the risk-neutral pricing of European options in Section \ref{PIsecH} along with the BSDE representation of the controlled value process. We show how to derive the optimal driver that generates the BSDE of the optimally controlled value processes, which gives us the pricing bounds for 
options under parameter uncertainty. To obtain actual values for the pricing bounds, we must resort to numerical solutions for the BSDE that governs the optimal value. In Appendix \ref{PIseqNum}, we detail some  simulation schemes for this purpose, and demonstrate the methods in a controlled setting to be able to compare and evaluate their performance. With a suggested numerical scheme in hand, we  proceed in Section \ref{PIsecE} to illustrate our method empirically on real-world market data. 
For a set of market quotes of European call options on the S\&P 500 index, we  
investigate how well the (numerically calculated) model bounds actually cover observed market prices. We also compare the results with the corresponding constant-parameter optimal price. For completion, we finally treat the general multi-asset case of a Markovian stochastic volatility model with jumps in Section \ref{PIsecG}. Section \ref{PIsecC} concludes.

\section{The Heston stochastic volatility model}\label{PIsecH}

To set the scene, we consider a financial market consisting of a risk-free money account and a risky asset over a fixed time period $[0,T]$. We assume the standard assumptions of a frictionless market: short selling is permitted and assets may be held in arbitrary amounts, there are no transaction costs and borrowing and lending are made at the same interest rate. The prices of the assets will be modelled as adapted stochastic processes on a filtered probability space, the notion of which will be formalised in the following section.

\subsection{European option pricing}
Let $(\Omega,\mathcal{F},\{\mathcal{F}_t\}_{t\geq0},P)$ be a filtered probability space where $\{\mathcal{F}_t\}_{t\geq0}$ is the natural filtration generated by two independent Wiener processes $W^1$ and $W^{2}$, augmented to satisfy the usual conditions of $P$-completeness and right continuity. We assume that the asset price $S$ and variance $V$ follow the model by \cite{heston1993closed}, with real-world dynamics (under the objective probability measure $P$)  given by
\begin{equation*}
\begin{aligned}
dS_t &= \mu(V_t) S_t dt + \sqrt{V_t}S_t  (\rho dW^1_t + \sqrt{1-\rho^2}dW^{2}_t),\\
dV_t &= \kappa(\theta-V_t)dt + \sigma\sqrt{V_t}dW^1_t,
\end{aligned}
\end{equation*} 
for nonnegative constants $\kappa$, $\theta$, $\sigma$ and instantaneous correlation $\rho\in(-1,1)$. The variance process thus follows a square root process\footnote{Also know as a CIR process from its use as a model for short-term interest rates by \cite{cox1985theory}. The square root process goes back to \cite{feller1951two}.} and it is bounded below by zero. If Feller's condition is satisfied, $2\kappa\theta\geq\sigma^2$,  the boundary cannot be achieved. Furthermore, the relative rate of return $\mu$ is taken to be a deterministic function of the variance. In addition to the risky asset, the market contains a risk-free money account which value processes is denoted $B$. The money account pays a constant rate of return $r$, which means that $B$ obeys the deterministic dynamics $dB_t = rB_tdt$.

The market price of risk processes $(\gamma^1,\gamma^2)$ associated with $W^1$ and $W^2$ are assumed to be specified such that
\begin{equation}\label{PIeqn29}
\frac{\mu(V)-r}{\sqrt{V}} = \left(\rho\gamma^1 + \sqrt{1-\rho^2}\gamma^2   \right)
\end{equation}
and as suggested by Heston, we let $\gamma^1 \equiv \lambda\sqrt{V}$ for some constant $\lambda$.\footnote{Heston motivates this choice  from the model of \cite{breeden1979intertemporal} under the assumption that the equilibrium consumption process also follows a square-root process; the risk premium is then  proportional to  variance. Aggregate risk preferences aside, a consequence is that the pricing equation \eqref{PIeqn31} conveniently  allows for Heston's pricing formula.    }
We then have that the stochastic exponential of  
$-(\gamma^1,\gamma^2)\bullet (W^1,W^2)$ is given by\footnote{We use $\bullet$ to denote the stochastic integral of $d$-dimensional processes: $H\bullet M =\sum_{i=1}^d \int_0^. H^i_tdM^i_t$ for $H,M$ taking values in $\R^d$. }
\begin{equation*}
\mathcal{E}(-\gamma\bullet W) = \exp\left(-\int_0^.\lambda\sqrt{V_s}dW^1_s -\int_0^.\gamma^2_sdW^{2}_s  - \frac{1}{2}\int_0^.(\lambda^2V_s + (\gamma^2_s)^2)ds \right)
\end{equation*}
and if we define the measure $Q$ on $\mathcal{F}_T$ for a fixed deterministic time $T$ by
\begin{equation*}
\frac{dQ}{dP} = \mathcal{E}(-\gamma\bullet W)_T
\end{equation*}
we have that $Q$ is equivalent to $P$ (provided the stochastic exponential is a martingale, i.e. $\E[\mathcal{E}(-\gamma\bullet W)_t]=1$ for all $t\in[0,T]$, for which Novikov's and Kazamaki's conditions are sufficient.  \cite{wong2006changes} express this explicitly in terms of the parameters). Further, by the Girsanov theorem, $\{\tilde{W}^1_t\}_{t\in[0,T]}$ and $\{\tilde{W}^{2}_t\}_{t\in[0,T]}$ defined by
\begin{equation*}
\begin{aligned}
d\tilde{W}^1_t &= \lambda\sqrt{V_t}dt + dW_t^1, \\
d\tilde{W}^{2}_t &= \gamma^2_tdt + dW^{2}_t,
\end{aligned}
\end{equation*} 
are independent Wiener processes under $Q$. By virtue of equation (\ref{PIeqn29}), this gives the $Q$-dynamics of the model
\begin{equation}\label{PIeqn13} 
\begin{aligned}
dS_t &= r S_t dt + \sqrt{V_t}S_t(\rho d\tilde{W}^1_t + \sqrt{1-\rho^2}d\tilde{W}^{2}_t) ,\\
dV_t &= \left(\kappa\theta - \left[\kappa+\sigma\lambda\right]V_t   \right)dt
+ \sigma\sqrt{V_t} d\tilde{W}^1_t,
\end{aligned}
\end{equation} 
 for $t\in[0,T]$ and we note that the variance dynamics are {form invariant} under the measure change: $V$ also follows a square root process under $Q$  with ``risk-neutral'' parameters $\tilde{\kappa},\tilde{\theta},\sigma$ where
\begin{equation*}
\tilde{\kappa} = \kappa+\sigma\lambda\,\,\,\,\,\text{and}\,\,\,\,\,\tilde{\theta}  = \frac{\kappa\theta}{\kappa+\sigma\lambda}.
\end{equation*} 
We also see that the discounted asset price $B^{-1}S$ will be a $Q$-martingale (i.e. $Q$ is an equivalent martingale measure) such that the financial market model $(B,S)$ is arbitrage-free. However, as $(\gamma^1,\gamma^2)$ may be arbitrarily chosen as long as (\ref{PIeqn29}) is satisfied, the model is incomplete. This means that $\lambda$ could be determined by a single exogenously given asset (with a volatility dependent price) to complete the market, and $\gamma^2$ is uniquely determined by equation (\ref{PIeqn29}). Any other contingent claim will then be uniquely priced.

For a European option with payoff $g(S_T)$ at maturity $T$, we have that the $\mathcal{C}^{1,2}$ function $D(t,s,v)$ of the pricing rule $D_t = D(t,S_t,V_t)$, $t\in[0,T]$, for the option satisfies the following partial differential equation
{\small
\vspace{5pt}
\begin{equation}\label{PIeqn31}
 \frac{\partial D}{\partial t} +rs\frac{\partial D}{\partial s}+\left\{ \kappa\theta - v(\kappa+\sigma\lambda) \right\}\frac{\partial D}{\partial v}
+\frac{1}{2}s^2v  \frac{\partial^2 D}{\partial s^2} + \rho\sigma vs \frac{\partial^2 D}{\partial v\partial s} + \frac{1}{2}\sigma^2v \frac{\partial^2 D}{\partial v^2} = rD,
\vspace{5pt}
\end{equation}
}with terminal condition $D(T,s,v)=g(s)$. Notice that the expression in curly brackets can be equivalently written $\tilde{\kappa}(\tilde{\theta}-v)$ with the risk-neutral specification of the parameters. Equivalently, by Feynman--Kac, this is to say that we have the usual risk-neutral pricing formula
\begin{equation*}
D(t,s,v) = \E^Q\left[\left. e^{-r(T-t)}g(S_T)\right| (S_t,V_t)=(s,v) \right]
\end{equation*}
where $(S,V)$ follows the $Q$-dynamics with initial value $(S_t,V_t)=(s,v)$ 
at the initial time $t\in[0,T]$.

The pricing equation (\ref{PIeqn31}) is the same as in Heston's original paper if we let $\lambda_v = \sigma\lambda$ and $\lambda(t,S_t,V_t)=\lambda_v V_t$  the price of volatility risk used in his exposition. The equation is solved by considering the Fourier transform of the price and the resulting ``semi-closed'' pricing formula is obtained by the inverse transform. In practice, however, the inverse transform has to be calculated by numerical integration methods. 


\subsection{Conservative pricing under parameter uncertainty}
Heston's model (and any other stochastic volatility model) is fundamentally a model for the underlying financial market even if it is predominantly used for option pricing purposes. The pricing measure is often taken as being fixed for convenience, for instance through model-to-market calibration of option prices, and the connection to the objective measure is not important for the analysis; and hence not necessarily made explicit. 

Although we are dealing with option pricing as well, we take a slightly converse approach in the case when the pricing measure inherits its uncertainty from the objective measure. Here we infer uncertainty of pricing parameters from statistical estimation of objective parameters, and the relation between the measures will thus play an integral role. On the other hand when uncertainty is deduced from a calibration method of pricing parameters directly, there is no need to make an explicit connection between the measures. We will handle both cases simultaneously and
for this purpose, we assume a pricing measure $Q$ to be given momentarily just to be able to replace it with another pricing measure that is subject to uncertainty.

\vspace{10pt}

\noindent To this end, we introduce \text{parameter uncertainty} in our model by modifying our reference measure with the effect of a control that governs the \text{parameter processes}. Namely, we replace the risk-neutral measure $Q$ with an equivalent measure $Q^u$ under which we have the controlled dynamics
\begin{equation}\label{PIeqn35} 
\begin{aligned}
dS_t &= \ru S_t dt + \sqrt{V_t}S_t(\rho d{W}^{u1}_t  + \sqrt{1-\rho^2}dW^{u2}_t) ,\\
dV_t &= \ku\left(\tu-V_t   \right)dt
+ \sigma\sqrt{V_t}d{W}^{u1}_t,
\end{aligned}
\end{equation} 
for $t\in[0,T]$. The control process $\{u_t\}_{t\geq0}$ 
is an $\mathcal{F}_t$-predictable process that takes values in a compact set $U\subset\R^3$, which we will call the \text{parameter uncertainty set}. We write $\mathcal{U}$ for the space of admissible control processes (that is,  predictable processes taking values in $U$ with sufficient integrability) and under $Q^u$, we have that ${W}^{u1}$ and ${W}^{u2}$ are independent Wiener processes, as will be explained in a moment. The control process realises its paths stochastically, and we simply do not know beforehand which $\{u_t\}_{t\geq0} \in \mathcal{U}$  will be governing (\ref{PIeqn35}): the uncertainty is tantamount to this choice.

Furthermore, we denote the components of the control $\{u_t\}_{t\geq0} = \{r_t,\kappa_t,\theta_t\}_{t\geq0}$ and let the controlled drift-functions of (\ref{PIeqn35}), all $f:U\rightarrow\R^+$, be defined as
\begin{equation}\label{PIeqn98} 
\begin{aligned}
\ru = r_t,\,\,\,\,\,\text{}\,\,\,\,\,
\ku = \kappa_t+\sigma\lambda \,\,\,\,\,\text{and}\,\,\,\,\,
\tu = \frac{\kappa_t\theta_t}{\kappa_t+\sigma\lambda}.
\end{aligned}
\end{equation} 
Notice that this specification of the controlled drift relies on the premise that the $Q$-parameters $r,\tilde{\kappa},\tilde{\theta}$ are subject to parameter uncertainty by their replacement with $r^u,\kappa^u,\theta^u$. 
The uncertainty is in turn taken to be inferred from statistical estimation of the objective $P$-parameters, represented by $(r_t,\kappa_t,\theta_r)\in U$ where $U$ is the statistical uncertainty set, 
and transferred to the pricing parameters by the map 
\begin{equation*}
U\ni u_t \mapsto \left(\ru,\ku,\tu\right)  \in U^\lambda
\end{equation*}
as given by (\ref{PIeqn98}). Here $U^\lambda$ is the uncertainty set for the controlled parameters, induced by the same mapping. The parameter $\lambda$ associated with $Q$  thus plays an instrumental role in facilitating the uncertainty transfer and it determines the set $U^\lambda$ where our uncertain price-parameters live. In practice, we forcefully set $\lambda=0$ to obtain that the uncertainty in price-parameters is exactly that of the uncertainty in estimated real-world parameters, i.e. $U^\lambda\equiv U$. However, note that this does not imply $P\equiv Q$ nor $\mu=r$, cf. equation (\ref{PIeqn29}). 

Similarly, when a calibration approach is employed to give pricing parameters and an associated uncertainty set directly, the identities 
\begin{equation}\label{PIeqn98aa} 
\begin{aligned}
\ru = r_t,\,\,\,\,\,\text{}\,\,\,\,\,
\ku = \kappa_t, \,\,\,\,\,\text{}\,\,\,\,\,
\tu = \theta_t
\end{aligned}
\end{equation} 
will facilitate the replacement of $r,\tilde{\kappa},\tilde{\theta}$ with the control $(r_t,\kappa_t,\theta_t)\in U$, now representing uncertain pricing parameters restricted to lie within a risk-neutral uncertainty set $U$.\footnote{Here we could be a bit more finical on notation, for instance with $(r_t,\tilde{\kappa}_t,\tilde{\theta}_t)\in \tilde{U}$ representing the pricing uncertainty deduced from calibration. For brevity, we refrain from such a notional distinction.}

\vspace{10pt}

\noindent With the controlled dynamics of $(S,V)$ representing the model under influence of parameter uncertainty, we proceed to define what we mean with the upper and lower boundary for the price of a European option. Namely, for an option written on $S$ with terminal payoff at time $T$ given by a square-integrable $\mathcal{F}_T$-measurable random variable $G$,  we will take the most conservative prices from valuation under the controlled pricing measures (i.e. under parameter uncertainty) as given by the control problems 
\begin{equation}\label{PIeqnCP} 
\begin{aligned}
D^-_t= \essinf_{\{u_t\}\in\mathcal{U}} \E_u\left[\left. e^{-\int_t^Tr_sds}G\right| \mathcal{F}_t  \right] \,\,\,\,\, \text{and}\,\,\,\,\,
D^+_t=\esssup_{\{u_t\}\in\mathcal{U}} \E_u\left[\left.e^{-\int_t^Tr_sds}G\right| \mathcal{F}_t  \right]
\end{aligned}
\end{equation}
for $t\in[0,T]$, where $\E_u(\cdot|\mathcal{F})$ denotes the conditional expectation under $Q^u$. In a sense, we thus consider the super-replication costs of selling a long/short position in the option when the uncertain parameters evolve stochastically in the uncertainty set, in an optimal way.\footnote{This draws on the interpretation that $\{Q^u:u\in\mathcal{U}\}$ is the set of equivalent martingale measures of an incomplete market model, such that the most conservative risk-neutral price of an option equals  the super-replication cost of a short position in the same: with $\Pi_t(G)=\inf_{\phi}\{\tilde{V}_t(\phi):V_T(\phi)\geq G,\,\text{a.s.}  \}$ being the discounted portfolio value of the (cheapest) admissible strategy $\phi$ that super-replicates $G$, then $\Pi_t(G)=\esssup_{u\in\mathcal{U}}\E_u[\tilde{G}|\mathcal{F}_t]$ and the supremum is attained.  See for instance \cite{contfinancial}, Section 10.2.}

In order to find a pricing PDE that corresponds to equation (\ref{PIeqn31}) of the previous section, we henceforth consider payoffs given by $G=g(S_T)$ for some non-negative function $g$. 
Due to the Markovian structure of the problem, we then have that the optimal value processes will be functions of the current asset price and variance state
\begin{equation*}
\begin{aligned}
D^-_t &= D^-(t,S_t,V_t), \\
D^+_t &= D^+(t,S_t,V_t), 
\end{aligned}
\end{equation*}
for some continuous functions $D^\pm:[0,T]\times\R^+\times\R^+\rightarrow\R$. As we will see later, these functions will satisfy a semilinear version of the standard pricing equation for European options. However, before we arrive at more precise expressions for the optimally controlled value processes (and their generating functions) we take one step backwards: we will first consider the value process for a fixed control 
and its link to a backward stochastic differential equation. Following the approach due to \cite{quenez1997stochastic} as outlined in \cite{cohen2015stochastic}, we then consider the optimally controlled value process as the solution to a closely related BSDE.

\vspace{10pt}

\noindent {With the intention of finding the pricing-boundaries  through a dual formulation with BSDEs, we begin by considering the following representation from Girsanov's theorem  }
\begin{equation}\label{hEj}
\begin{aligned}
d{W}^{1u}_t &= d\tilde{W}^1_t - \alpha_1(t,S_t,V_t)dt, \\
d{W}^{2u}_t &= d\tilde{W}^2_t - \alpha_2(t,S_t,V_t)dt.
\end{aligned}
\end{equation}
For its kernel $\alpha = (\alpha_1,\alpha_1)^{\top}$, straightforward algebra then shows that 
\begin{equation}\label{PIeqn11}
\alpha(S_t,V_t,u_t) = \frac{1}{\sigma\sqrt{V_t}}
  \begin{bmatrix}
	\ku\tu-\tilde{\kappa}\tilde{\theta}-(\ku-\tilde{\kappa})V_t \\
	\frac{-\rho\left(\ku\tu-\tilde{\kappa}\tilde{\theta}-(\ku-\tilde{\kappa})V_t\right)+\sigma(r_t-r)}{\sqrt{1-\rho^2}}
  \end{bmatrix}
\end{equation}
ties together the dynamics \eqref{PIeqn35} and \eqref{PIeqn13} associated with $W^u$ and $\tilde{W}$, respectively.  Formally, if the stochastic exponential of the process $\alpha(S,V,u)^{\top}\bullet(\tilde{W}^1,\tilde{W}^2)$ defines the measure change $Q\rightarrow Q^u$ on $\mathcal{F}_T$,
\begin{equation*}
\frac{dQ^u}{dQ} = \mathcal{E}\left(\int_0^.\alpha_1(S_t,V_t,u_t)d\tilde{W}^1_t+\int_0^.\alpha_2(S_t,V_t,u_t)d\tilde{W}^2_t\right)_T,
\end{equation*}
(provided $\mathcal{E}(\alpha^{\top}\bullet\tilde{W})$ is a martingale), then $W^{1u}$ and $W^{2u}$ defined in \eqref{hEj}  are two independent Wiener processes under $Q^u$ by Girsanov's theorem.

Next, we define the linear driver function $f:(0,\infty)\times(0,\infty)\times\R\times\R^{1\times2}\times U\rightarrow\R$ as
\begin{equation}\label{PIeqn10}
f(s,v,y,z,u) = \varrho(s,v,u)y + z\alpha(s,v,u)
\end{equation}
where $\varrho(s,v,u) \equiv -r$, that is, the (negative) first component of the control which represents the risk-free interest rate. 
With the driver defined by (\ref{PIeqn11})-(\ref{PIeqn10}), we then obtain the following representation of the 
expected value under $Q^u$: for a  given, fixed control process $u=\{u_t\}_{t\geq0}\in\mathcal{U}$, the controlled value process  given by
\begin{equation*}
J_t(u) = \E_u\left[\left. e^{-\int_t^Tr_sds}g(S_T)\right| \mathcal{F}_t  \right],\,\,\,\,\,t\in[0,T],
\end{equation*}
is the {unique} solution to the linear Markovian backward stochastic differential equation
\begin{equation}\label{PIeqn1}
\begin{aligned}
dJ_t(u) &= -f(S_t,V_t,J_t(u),Z_t,u_t)dt+Z_td\tilde{W}_t, \\
J_T(u) &= g(S_T),
\end{aligned}
\end{equation}
where $Z=(Z^1,Z^2)^\top$---the martingale representation part of $J(u)$---is a process taking values in $\R^{1\times 2}$ (being a part of the solution to the BSDE). To see this, consider the process $J(u)$ that solves (\ref{PIeqn1}) and let $\mathcal{E}(\Gamma)$ be the stochastic exponential of
\begin{equation*}
\Gamma = \int_0^.-r_tdt + \int_0^.\alpha(S_t,V_t,u_t)^{\top}d\tilde{W}_t.
\end{equation*}
Apply It\^{o}'s product rule to $\mathcal{E}(\Gamma)J(u)$ to obtain
\begin{equation*}
d\left(\mathcal{E}(\Gamma)_tJ_t(u)\right) = \mathcal{E}(\Gamma)_t\left(Z_t+J_t(u)\alpha(S_t,V_t,u_t)^{\top}\right)d\tilde{W}_t
\end{equation*}
\noindent and thus, since $\mathcal{E}(\Gamma)J(u)$ is a martingale under $Q$, we have
\begin{equation*}
\begin{aligned}
J_t(u) &= \frac{1}{\mathcal{E}(\Gamma)_t}\E^Q\left[\left.\mathcal{E}(\Gamma)_T\,g(S_T)    \right|\mathcal{F}_t\right] \\
&=e^{\int_0^tr_sds}\frac{1}{\mathcal{E}(\alpha^{\top}\bullet\tilde{W})_t}\E^Q\left[\left.e^{-\int_0^Tr_sds}\mathcal{E}(\alpha^{\top}\bullet\tilde{W})_T\,g(S_T)    \right|\mathcal{F}_t\right] \\
&=\E_u\left[\left.e^{-\int_t^Tr_sds}g(S_T)    \right|\mathcal{F}_t\right]
\end{aligned}
\end{equation*}
as $\mathcal{E}(\alpha^{\top}\bullet\tilde{W})$ is the density for the measure change $Q\rightarrow Q^u$.

\vspace{10pt}

\noindent The BSDE (\ref{PIeqn1}) governs the value process under the impact of a fixed parameter control process that evolves in the uncertainty set $U$. To obtain the lowest (highest) value scenario, the value process is to be minimised (maximised) over all admissible controls in $\mathcal{U}$, and as we detail in the following, this is done through {pointwise} optimisation with respect to $u\in U$ 
of the driver function for the value process. 
Hence, we define the following drivers optimised over the parameter uncertainty set
\begin{equation*}
\begin{aligned}
H^-(s,v,y,z) = \essinf_{u\in U} f(s,v,y,z,u) \quad \text{and}\quad
H^+(s,v,y,z) = \esssup_{u\in U} f(s,v,y,z,u),
\end{aligned}
\end{equation*}
where we note that as $U$ is compact, the infimum and supremum are both attained. We then have the following main result which is due to the {comparison principle for BSDEs}:
the lower/upper optimally controlled value processes 
\begin{equation*}
\begin{aligned}
D^-_t = \essinf_{u\in \mathcal{U}} J_t(u) \quad \text{and} \quad
D^+_t = \esssup_{u\in \mathcal{U}} J_t(u)
\end{aligned}
\end{equation*}
for $t\in[0,T]$, have cadlag modifications that are the {unique} solutions of the BSDEs 
\begin{equation}\label{PIeqn2}
\begin{aligned}
dD^\pm_t &= -H^\pm(S_t,V_t,D^\pm_t,Z_t)dt+Z_td\tilde{W}_t, \\
D^\pm_T &= g(S_T).
\end{aligned}
\end{equation}
\noindent In particular, the processes are equal to deterministic functions of $(t,S_t,V_t)$, that is, $D^\pm_t=D^{\pm}(t,S_t,V_t)$ for some continuous functions $D^{\pm}:[0,T]\times\R^+\times\R^+\rightarrow\R$. As the infimum (supremum) of $H$ is attained, we further have that there exists optimal controls $\{u^{\pm*}_t\}_{t\in[0,T]}\in\mathcal{U}$ which are  \textit{feedback} controls. This means that the processes
\begin{equation*}
u^{\pm*}_t = u^{\pm*}(t,S_t,V_t),\,\,\,\,\,t\in[0,T],
\end{equation*}
are the optimal controls among all predictable controls for some deterministic functions $u^{\pm*}:[0,T]\times\R^+\times\R^+\rightarrow U$. Finally, by the semilinear Feynman-Kac formula (provided  a solution exists), we have that $D^-(t,s,v)$ satisfies the following semilinear parabolic PDE
\begin{align}\label{PIeqn26}
& \frac{\partial D}{\partial t} +\frac{1}{2}s^2v  \frac{\partial^2 D}{\partial s^2} + \rho\sigma vs \frac{\partial^2 D}{\partial v\partial s} + \frac{1}{2}\sigma^2v \frac{\partial^2 D}{\partial v^2} \\
&+  \essinf_{(r,\kappa,\theta)\in U} \left\{-rD+rs \frac{\partial D}{\partial s}+\kappa^u(\kappa)\left(\theta^u(\theta)-v\right)\frac{\partial D}{\partial v} \right\}=0 \nonumber
\end{align}
with terminal value $D^-(T,s,v)=g(s)$. In the corresponding equation for $D^+(t,s,v)$ we have a supremum substituted for the infimum.

\vspace{10pt}
\noindent \textit{Proof:} For the first part of the result, since $H^-(s,v,y,z) \leq f(s,v,y,z,u)$ by definition, we have that the (unique) solution\footnote{As we assume the driver $f$ to be sufficiently integrable for the $J(u)$-BSDE to admit a unique solution (i.e. it is a stochastic Lipschitz driver) the integrability carries over to $H$ such that the $Y$-BSDE admits a unique solution as well.} $Y$ to the BSDE with data $(g(S_T),H^-)$ satisfies $Y_t\leq J_t(u)$ for all controls $u\in\mathcal{U}$ (up to indistinguishability). This is a consequence of the {comparison theorem for BSDEs} (\cite{Peng1992}, see also \cite{El1997}). 
Further, by {Filippov's implicit function theorem} (\cite{3}, see also \cite{McShane} and \cite{Filippov}), 
for each $\epsilon>0$ there exists a predictable control $u^\epsilon\in\mathcal{U}$ such that $f(s,v,y,z,u^\epsilon)\leq H^-(s,v,y,z)+\epsilon$. Since $Y_t+\epsilon(T-t)$ solves the BSDE with driver $H^-(s,v,y,z)+\epsilon$, the comparison theorem yields $J_t(u^\epsilon)\leq Y_t+\epsilon(T-t)$ (up to indistinguishability) and we have the inclusion
\begin{equation*}
Y_t\leq J_t(u^\epsilon) \leq Y_t + \epsilon(T-t).
\end{equation*}
Letting $\epsilon\rightarrow0$ we have that $Y_t = \essinf_{u\in\mathcal{U}} J_t(u)=D^-_t$ for every $t$ which is to say that $Y$ is a version of the optimal value process. That $D^-_t$ can be written as a continuous function of $(t,S_t,V_t)$ is due to the fact that (\ref{PIeqn2}) is a Markovian BSDE. Further, as we have that the optimal control is attainable, Filippov's theorem gives that it is a function of $(t,S_t,V_t,D^-_t,Z_t)$ where $Z_t=z(t,S_t,V_t)$---due to the Markovian BSDE\footnote{The function for the martingale representation $Z$ is obtained explicitly by applying It\^{o}'s lemma to $D_t=D(t,S_t,V_t)$ and using the semilinear pricing PDE (\ref{PIeqn26}), which gives
\begin{equation*}
dD(t,S_t,V_t) = -H(S_t,V_t,D_t,Z_t)dt + \partial_xD(t,S_t,V_t) \sigma(S_t,V_t)d\tilde{W}_t
\end{equation*}
where $\partial_xf\equiv(\partial_sf,\partial_vf)$ and $\sigma(s,v)$ should be understood as the diffusion matrix of (\ref{PIeqn13}). Hence, by uniqueness of the BSDE solution, $z(t,s,v) \equiv \partial_xD(t,s,v) \sigma(s,v)$ is the deterministic generating function for $Z$.\label{PIfn1}}---and we have the result that $u^{-*}_t = u^{-*}(t,S_t,V_t)$ for a deterministic function. 
\hfill{$\square$}

\vspace{10pt}
\noindent To obtain an expression for the optimised driver $H^{\pm}$, we note that the driver of the value function is conveniently expressed in terms of divergence of the controlled drift from the original parameters; by rearrangement of (\ref{PIeqn10})
{\footnotesize
\begin{align}\label{PIeqn12bb}
f(S_t,V_t,Y_t,Z_t,u_t) &= (r_t-r)\left( \frac{Z^2_t}{\sqrt{1-\rho^2}\sqrt{V_t}}-Y_t \right) +(\ku-\tilde{\kappa})\left( \frac{-Z^1_t\sqrt{V_t}}{\sigma}+\frac{\rho Z^2_t\sqrt{V_t}}{\sigma\sqrt{1-\rho^2}}   \right)\\ 
& + (\ku\tu - \tilde{\kappa}\tilde{\theta})\left(\frac{Z_t^1}{\sigma\sqrt{V_t}}-\frac{\rho Z^2_t}{\sigma\sqrt{1-\rho^2}\sqrt{V_t}} \right)- rY_t. 
\end{align}
}Alternatively, this can be expressed as
{\footnotesize
\begin{align}\label{PIeqn12}
f(S_t,V_t,Y_t,Z_t,u_t) &= (r_t-r)\left( \frac{Z^2_t}{\sqrt{1-\rho^2}\sqrt{V_t}}-Y_t \right) +(\kappa_t-\kappa)\left( \frac{-Z^1_t\sqrt{V_t}}{\sigma}+\frac{\rho Z^2_t\sqrt{V_t}}{\sigma\sqrt{1-\rho^2}}   \right)\\ 
& + (\kappa_t\theta_t - \kappa\theta)\left(\frac{Z_t^1}{\sigma\sqrt{V_t}}-\frac{\rho Z^2_t}{\sigma\sqrt{1-\rho^2}\sqrt{V_t}} \right)- rY_t 
\end{align}
}since $\ku-\tilde{\kappa}=\kappa_t-\kappa$ and $\ku\tu-\tilde{\kappa}\tilde{\theta}=\kappa_t\theta_t-\kappa\theta$, which is due to the linear form of the drift (and the simple form of the parameter change under $P\rightarrow Q$, regardless of the value of $\lambda$). 
If we let $\beta_t\equiv\kappa_t\theta_t$ and use the parametrisation $(r_t,\kappa_t,\beta_t)\mapsto(r_t,\kappa_t,\theta_t)$, we thus have that the driver is a linear function of the divergence 
\[
\tilde{u}_t=(r_t-r,\kappa_t-\kappa,\beta_t-\beta).
\]
Hence, the optimal drivers $H^{\pm}$ are obtained by minimising/maximising a linear objective subject to the constraint given by the compact uncertainty set $U$ 
---equation (\ref{PIeqn12}) in case of statistical uncertainty for $P$-parameters, or similarly, (\ref{PIeqn12bb}) with the identities (\ref{PIeqn98aa}) when dealing with $Q$-parameters under uncertainty deduced from calibration. 

To this end, we consider elliptical uncertainty sets as given by the quadratic form
\begin{equation*}
U=\left\{u:  \tilde{u}^\top\Sigma^{-1}\tilde{u} \leq \chi  \right\}
\end{equation*}
for some 
positive semi-definite matrix $\Sigma$ and positive constant $\chi$. In particular, from statistical inference, 
we have that the $1-\alpha$ confidence ellipse
\begin{equation}\label{PIeqn99}
\tilde{u}^\top\Sigma_{r,\kappa,\beta}^{-1}\tilde{u} \leq \chi^2_3(1-\alpha)
\end{equation}
represents $u\in U$ (for a significance level $\alpha$) where $\Sigma_{r,\kappa,\beta}$ is the covariance matrix of the parameters and $\chi^2_3(1-\alpha)$ is the quantile of the chi-squared distribution with three degrees of freedom (see further Section \ref{PIsecEst}). {The formal justification for elliptical uncertainty sets is that} \eqref{PIeqn99} {is a level set of the asymptotic likelihood, due to large-sample normality of the maximum likelihood estimator of the parameters. } This is the same form of uncertainty as given by the confidence region under Wald's hypothesis test, and approximates the optimal confidence set, by Neyman--Pearson lemma  (see \cite{lehmann2006testing}).

As $\tilde{u}\mapsto f(\tilde{u})$ is linear, it has no internal stationary points and the quadratic problems
\begin{equation*}
\begin{aligned}
&H^- = \inf f(\tilde{u})\,\,\,\,\,\text{and} \,\,\,\,\,H^+ = \sup f(\tilde{u})\\
&\text{subject to } \tilde{u}^\top\Sigma^{-1}\tilde{u} = \chi
\end{aligned}
\end{equation*}
give the optimised drivers. The solutions are (obtained e.g. by a Lagrange multiplier)
\begin{equation}\label{PIeqn19}
\begin{aligned}
&H^\pm(S_t,V_t,Z_t,Y_t) = \pm\sqrt{\chi\, n_t^{\top}\Sigma^{\top}n_t} - rY_t\\
&\tilde{u}^{\pm}(S_t,V_t,Z_t,Y_t) = \pm\sqrt{\frac{{\chi}}{{n_t^{\top}}\Sigma^{\top}n_t}}\Sigma \,n_t
\end{aligned}
\end{equation}
where $n_t$ is the $3\times1$ vector of coefficients to the parameter deviances of equation (\ref{PIeqn12}) given by
\begin{equation*}
n_t = \left[ \left( \frac{Z^2_t}{\sqrt{1-\rho^2}\sqrt{V_t}}-Y_t \right),\left( \frac{-Z^1_t\sqrt{V_t}}{\sigma}+\frac{\rho Z^2_t\sqrt{V_t}}{\sigma\sqrt{1-\rho^2}}   \right),  \left(\frac{Z_t^1}{\sigma\sqrt{V_t}}-\frac{\rho Z^2_t}{\sigma\sqrt{1-\rho^2}\sqrt{V_t}} \right)  \right]^\top.
\end{equation*}

The optimal drivers in (\ref{PIeqn19}) conclude our analysis since we now have an explicit  form for the stochastic differential equation (\ref{PIeqn2}) that describes the evolution of the pricing boundaries. Before we proceed to obtaining approximative solutions of these equation by numerical methods, a few remarks are in order. Firstly, the approach applies unchanged to a portfolio of options with time-$T$ terminal payoff $\sum_ig_i(S_T)$. Due to the non-linearity of the pricing boundaries (\ref{PIeqnCP}), we further have $D^+(\sum w_ig_i(S_T))\leq \sum w_i D^+(g_i(S_T))$ for weights $\sum w_i=1$, such that the super-replication cost for individual hedging might be lowered by hedging the portfolio as a whole. Secondly, for a general payoff represented by $G\in\mathcal{F}_T$, for instance a path-dependent European options on $S$, we have a value process equation corresponding to (\ref{PIeqn2}) with terminal condition $D^{\pm}_T=G$. However, this problem do no longer yield a Markovian structure, and we do not have $D^{\pm}$ (nor $Z$) being generated by deterministic functions, neither do the numerical methods of Appendix \ref{PIseqNum} apply. Thirdly, we deliberately impose parameter uncertainty by replacing $Q\rightarrow Q^u$ in contrast to replacing $P\rightarrow Q^u$ directly (which would yield the same form of the effect $\alpha(t,S_t,V_t)$ that governs the measure change, but with $r$ replaced by $\mu$ in (\ref{PIeqn11})). The reason is that the governing BSDEs (\ref{PIeqn2}) will have a terminal condition $g(S_T)$ where $S_T\sim Q$, for which we have accessible parameters (in particular, we may directly observe the $Q$-drift $r$ instead of estimating the $P$-drift $\mu$). 

We conclude this section with a  technical remark.

\begin{remark}
So far, we have not expressed any integrability conditions on the process $\alpha(S_t,V_t,u_t)$ in order to guarantee that (i) the density $dQ^u/dQ$ and (ii) the driver $f(S_t,V_t,Y_t,Z_t,u_t)$ are well defined, i.e. for the measure change $Q\rightarrow Q^u$ to be eligible and to certify that $f$ (and hence $H^\pm$) yields a BSDE which admits a unique solution. For this purpose, Novikov's condition
\begin{equation}\label{PIeqn5}
\E^Q\left[e^{\frac{1}{2}\int_0^T||\alpha(S_t,V_t,u_t)||^2dt}  \right] < \infty
\end{equation}
is sufficient for both (i) and (ii) since then we have that the driver is \text{stochastic Lipschitz} in $y$ and $z$ (note that $r_t$ is bounded in $U$), i.e.
\begin{equation*}
|f(S_t,V_t,y,z,u_t) - f(S_t,V_t,y',z',u_t) | \leq ||\alpha(S_t,V_t,u_t)||\,\left(|y-y'| + ||z-z'||\right)
\end{equation*}
where $||\alpha(S_t,V_t,u_t)||$ is predictable and such that (\ref{PIeqn5}) holds. With a stochastic Lipschitz driver, the concerned BSDE admits a unique solution which is bounded if the terminal condition $g(S_T)$ is bounded (see e.g. \cite{cohen2015stochastic}, Appendix A.9.2).

For Novikov's condition in (\ref{PIeqn5}) we note that the integrand of the exponent can be written
\begin{equation*}
||\alpha(S_t,V_t,u_t)||^2 = a(u_t)V_t + b(u_t) \frac{1}{V_t} + c(u_t) \\
\end{equation*}
with
\begin{equation*}
\begin{aligned}
a(u_t) &= \frac{(\kappa_t-{\kappa})^2}{\sigma^2(1-\rho^2)}\\
b(u_t) &= \frac{\sigma^2(r_t-r)^2 + (\kappa_t\theta_t-{\kappa}{\theta})^2 + 2\rho\sigma(r-r_t)(\kappa_t\theta_t-{\kappa}{\theta})}{\sigma^2(1-\rho^2)}\\
c(u_t) &= -2\frac{\left(\sigma\rho(r-r_t)+\kappa_t\theta_t-{\kappa}{\theta}\right)(\kappa_t-{\kappa})}{\sigma^2(1-\rho^2)}
\end{aligned}
\end{equation*}
such that for the expectation in (\ref{PIeqn5}) we have
\begin{equation}\label{PIeqn6}
\E^Q\left[e^{\frac{1}{2}\int_0^T  a(u_t)V_t   dt} e^{\frac{1}{2}\int_0^T  b(u_t)\frac{1}{V_t}   dt} e^{\frac{1}{2}\int_0^T   c(u_t)  dt}  \right] \leq k \sqrt{ \E^Q\left[e^{\int_0^T  a(u_t)V_t   dt}\right] \E^Q\left[e^{\int_0^T  b(u_t)\frac{1}{V_t}   dt}\right]}
\end{equation}
since $ e^{\frac{1}{2}\int_0^T   c(u_t)  dt}$ is bounded by a constant $k$ (for $u_t\in U$) and where we have used the Cauchy-Schwarz inequality. If we begin with the first expectation on the right hand side of (\ref{PIeqn6}) we have $a(u_t)\leq \bar{a}$ for a constant $\bar{a}$. As the Laplace transform of the integrated CIR process\footnote{The Laplace transform of the integrated variance $\E[\exp(-\beta\int_0^TV_tdt)]$ goes back to \cite{cox1985theory} and is well defined for $-\beta\leq\kappa^2/(2\sigma^2)$, see also \cite{carr2003stochastic}.} is finite for $\bar{a}\leq\tilde{\kappa}^2/(2\sigma^2)$, we end up with the condition
\begin{equation}\label{PIeqn7}
|\kappa_t-{\kappa}| \leq \tilde{\kappa}\frac{\sqrt{1-\rho^2}}{\sqrt{2}}.
\end{equation}
For the second expectation of (\ref{PIeqn6}), we use that $b(u_t)\leq\bar{b}$ for a constant $\bar{b}$ and that the Laplace transform of the integrated inverse-CIR process\footnote{\cite{carr2007new} gives an expression for the joint transform of the log-price and integrated variance of a 3-over-2 process. Applying It\^{o}'s formula to $1/V_t$ we find that the inverse-CIR $(\kappa,\theta,\sigma)$ process is a 3-over-2 process with parameters $(\hat{\kappa}\equiv\kappa\theta-\sigma^2,\hat{\theta}\equiv\kappa/(\kappa\theta-\sigma^2),\hat{\sigma}\equiv-\sigma)$. Using their transform, provided $\hat{\kappa}>-\hat{\sigma}^2/2$, 
\begin{equation*}
\E\left[e^{-\lambda\int_0^T \frac{1}{V_t}dt} \right] = \frac{\Gamma(\gamma-\alpha)}{\Gamma(\gamma)}\left(  \frac{2}{\hat{\sigma}^2y(0,1/V_0)} \right)^\alpha M\left(\alpha,\gamma,-\frac{2}{\hat{\sigma}^2 y(0,1/V_0)} \right) 
\end{equation*}
where
\begin{eqnarray*}
& y(t,x) \equiv x(e^{\hat{\kappa}\hat{\theta} (T-t)}-1)/(\hat{\kappa}\hat{\theta})=  x(e^{\kappa (T-t)}-1)/\kappa  \\
&\alpha \equiv -(1/2+\hat{\kappa}/\sigma^2) + \sqrt{(1/2+\hat{\kappa}/\sigma^2)^2 + 2\lambda/\sigma^2} \\
& \gamma \equiv 2(\alpha+1+\hat{\kappa}/\sigma^2) = 1+2\sqrt{(1/2+\hat{\kappa}/\sigma^2)^2 + 2\lambda/\sigma^2}
\end{eqnarray*}
and $M$ is the confluent hypergeometric function. From this, we see that
\begin{equation*}
\lambda\geq-\left(\frac{2\hat{\kappa}+\sigma^2}{2\sqrt{2}\sigma}  \right)^2=-\left(\frac{2{\kappa}\theta-\sigma^2}{2\sqrt{2}\sigma}  \right)^2
\end{equation*}
is a sufficient condition for the transform to being well defined.}
is finite for
\begin{equation}\label{PIeqn8}
\bar{b}\leq \left(\frac{2\tilde{\kappa}\tilde{\theta}-\sigma^2}{2\sqrt{2}\sigma}\right)^2.
\end{equation}
Rearranging this condition, we have that
\begin{equation*}
{\sigma^2(r_t-r)^2 + (\kappa_t\theta_t-{\kappa}{\theta})^2 + 2\rho\sigma(r-r_t)(\kappa_t\theta_t-{\kappa}{\theta})} \leq \frac{1-\rho^2}{2}\left({{\kappa}{\theta}-\sigma^2/2}\right)^2
\end{equation*}
together with (\ref{PIeqn7}) are sufficient conditions for (\ref{PIeqn5}) to hold. 
\end{remark}

\section{The empirical perspective}\label{PIsecE}
In this section we take a look at how the conservative  pricing method carries over to a set of real market data with historical prices of the S\&P 500 index. We perform an empirical experiment and here we have the following rationale as a base for our study. 

We let the statistical parameters as estimated from historical observations of the index price and variance represent the financial market model under the objective measure $P$. Hence, $(S,V)$ is assumed to evolve with $P$-dynamics according to Heston's model specified with the estimated parameters. We consider $(B,S)$ exclusively as the traded assets in the financial market model driven by two random sources, $W^1$ and $W^2$, and refrain from the assumption that there exists an additional, exogenously given, (volatility dependent) asset which would complete the model. On the other hand, we exclude arbitrage opportunities in the  model and affirm the existence of a space of risk-neutral measures: $Q$ exists (not necessarily unique) in a set $\mathcal{Q}$ of probability measures equivalent to $P$, such that the discounted asset price is a martingale under any measure in $\mathcal{Q}$. In our model context, this implies that $(S,V)$ will have the same diffusion matrix under every $Q\in\mathcal{Q}$ as given by the diffusion matrix of the $P$-dynamics. This follows from the notion that the quadratic variation (continuous part) of a semimartingale is invariant under equivalent probability measures on a complete filtered space. Further, by Girsanov's theorem, we have that the law of the driving random sources is invariant: they will be (independent) Wiener process under all equivalent measures in $\mathcal{Q}$.

With this in mind, we fix the diffusion matrix of $(S,V)$ to be as given by the estimated diffusion parameters from historical data. We then take the space of equivalent risk-neutral measures to be equal the space spanned by the controlled measure $Q^u$ over all admissible controls: $\mathcal{Q}\equiv\{Q^u:  u\in\mathcal{U}  \}$ where $\mathcal{U}$ represents the space of predictable control processes $u=\{u_t\}_{t\geq 0}$ that lives in the compact uncertainty set $U\subset\R^d$. In particular, we deduce the uncertainty set from inference of the statistical estimation problem and define $U$ to be represented by the elliptical confidence region 
as derived from the observed Fisher information (which asymptotically approximates the sampling covariance of our estimates). The question we ask is then if market option prices are covered by the pricing rules in $\mathcal{Q}$ as given by the corresponding model pricing-boundaries.

Since the volatility process of an asset is latent by nature it has to be measured with some method. In the following section we briefly present the realized volatility measure which gives a commonly used nonparametric estimator of the variance process. 
We then detail some estimation methods that we employ for point estimation of the model parameters and for drawing inference thereof. The empirical study based on S\&P 500 data finally follows.

\subsection{Measured variance and statistical inference}\label{PIsecEst}

\begin{figure}[t!]
\makebox[\textwidth][c]{  
\includegraphics[scale=0.44,trim=0 20 20 20,clip]{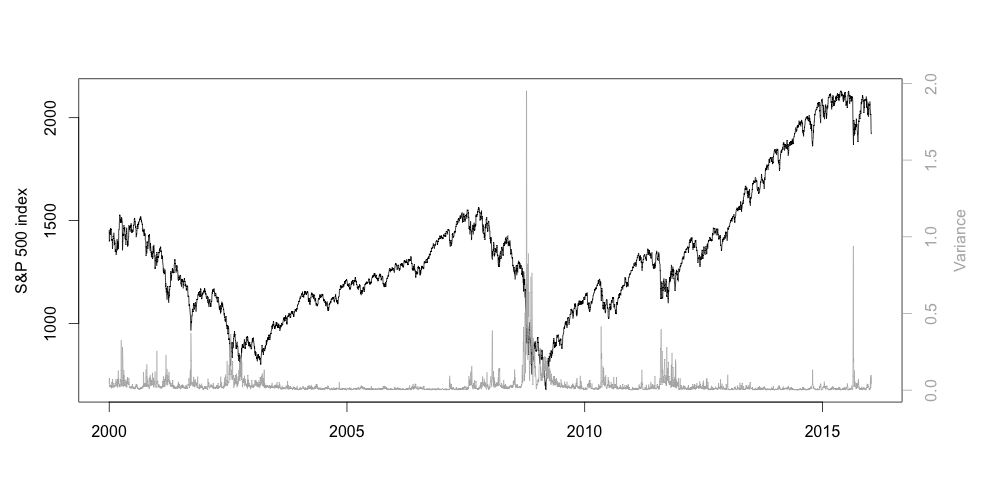}
}
\caption{Historical closing prices and realized variances of the S\&P 500 index, 4,035 daily observations from January 3\textsuperscript{rd}, 2000 to February 29\textsuperscript{th}, 2016. 
}
\label{PIfig6}
\end{figure}

We use the variance of the S\&P 500 price as estimated with the \textit{realised volatility measure} from high-frequency observations ($\sim$5min) of the index returns (see for instance \cite{andersen2009realized}). Hence, if $s=(s_{t_1},s_{t_2},\dots)$ is a time-series with daily prices, the realized variance measure is calculated as 
\begin{equation*}
RV([t_{i-1},t_{i}]) = \sum_{k:s_k,s_{k+1}\in[t_{i-1},t_{i}]} (y_{s_{k+1}}-y_{s_k})^2,
\end{equation*}
where $[t_{i-1},t_i]$ is the duration of the day 
and $s_k\in[t_{i-1},t_i]$ are intra-day time points over which  $y_{s_k}=\log(s_{s_k})-\log(s_{s_{k-1}})$ are the log-returns. The realized variance approximates the integrated variance: $RV([t_{i-1},t_{i}]) \Pconv \int_{t_{i-1}}^{t_i} V_s ds$, (for a continuous return process, the quadratic variation for a general semimartingale), and the measured variance at time $t_i$ is taken to be 
\begin{equation*}
v_{t_i} = \frac{1}{t_i-t_{i-1}}RV([t_{i-1},t_{i}]).
\end{equation*}
such that a time-series of daily variances $v=(v_{t_1},v_{t_2},\dots)$ is obtained. For convenience, we use precomputed variance estimates from the Oxford-Man Institute's realised library\footnote{The Realised Library version 0.2 by Heber, Gerd, Lunde, Shephard and Sheppard (2009), \url{http: //realized.oxford-man.ox.ac.uk}.} which are shown in 
Figure \ref{PIfig6} together with the closing price of the S\&P 500 index from the period January 3\textsuperscript{rd}, 2000 to February 29\textsuperscript{th}, 2016.

Next, we consider the estimation of the parameters $\Theta = (\kappa,\theta,\sigma)$ from $n+1$ observations $v=(v_0,\dots,v_n)$ of the variance. Here, $v_i$ is treated as the observed value of $V_{t_i}$ for a set of discrete time-points $(t_0,\dots,t_n)$ at which the observations are made, and we denote with $\Delta_i=t_{i+1}-t_i$ the length of the $i$\textsuperscript{th} time-interval between two consecutive observations. In general, inference on the parameters $\Theta$ of a Markov 
process with transition density $f(y;x,\delta,\Theta)$ for $V_{t+\delta}|V_t=x$ can be made with the likelihood function
\begin{equation*}
L_n(\Theta) = \prod_{i=0}^{n-1} f(v_{i+1};v_i,\Delta_i,\Theta)
\end{equation*}
and the maximising argument of the (log-) likelihood function is the maximum likelihood estimator $\hat{\Theta}$ of $\Theta$. If the transition density is unknown in closed form, or, as in the case for the square root process, of a kind that is impenetrable for optimisation (both analytically and by numerical schemes), one might consider alternatives based on suitable approximations of the likelihood. A direct way of doing so is to consider the time-discrete approximation ${V}^{\pi}$ of the process $V$ as given by a Euler-Maruyama scheme; here for the square root process\footnote{Note that \eqref{eqEMscheme} may generate negative outcomes of ${V}^{\pi}_{t+\delta}$ and is thus not suitable for simulation in its standard form. Alternative schemes are discussed in Appendix \ref{AppHe}. Here we use \eqref{eqEMscheme} for an approximative Gaussian likelihood---the Euler contrast \eqref{PIeqn9}---which is well defined. }
\begin{equation}\label{eqEMscheme}
{V}^{\pi}_{t+\delta} = {V}^{\pi}_t + \kappa(\theta-{V}^{\pi}_t)\delta + \sigma\sqrt{{V}^{\pi}_t}\left(W_{t+\delta}-W_{t}\right)
\end{equation}
which will give an approximative Gaussian log-likelihood function
\begin{equation} \label{PIeqn9}
l_n(\Theta) \equiv
\log L_n(\Theta) = -\frac{1}{2}\sum_{i=0}^{n-1} \frac{(v_{i+1}-v_i-\kappa(\theta-v_i)\Delta_i)^2}{\sigma^2v_i\Delta_i} + \log(2\pi\sigma^2v_i\Delta_i). 
\end{equation}
A function on the same form as above was considered for least-squares estimation of drift parameters in \cite{prakasa1983asymptotic}. For processes with ergodic property, \cite{kessler1997estimation} considered the joint estimation of drift and diffusion parameters with a Gaussian approximation to the transition density as (\ref{PIeqn9}) and showed that under general conditions, the estimator is asymptotically normal and efficient. Their approach addresses the case when the mean and variance of the transition density are unknown and uses approximations in their place. For the square root process, the explicit expressions
\begin{eqnarray*}
\begin{aligned}
\E[V_{t+\delta}|V_t=v] &= \theta+(v-\theta)e^{-\kappa\delta} \equiv \mu(v,\delta) , \\
\Var(V_{t+\delta}|V_t=v) & = {v \frac{\sigma^2}{\kappa}(e^{-\kappa\delta}-e^{-2\kappa\delta}) + \theta\frac{\sigma^2}{2\kappa}(1-e^{-\kappa\delta})^2} \equiv s^2(v,\delta),
\end{aligned}
\end{eqnarray*}
in place of the approximations $\mu(v,\delta) \approx v+\kappa(\theta-v)\delta$ and $s^2(v,\delta)\approx \sigma^2v\delta$ in (\ref{PIeqn9}) give that
\begin{equation}\label{eqExact}
l_n(\Theta) = -\frac{1}{2}\sum_{i=0}^{n-1} \frac{\left(v_{i+1}-\mu(v_i,\Delta_i)\right)^2}{s^2(v_i,\Delta_i)} + \log(2\pi s^2(v_i,\Delta_i))
\end{equation}
 forms an approximative Gaussian likelihood function with exact expressions for the conditional mean and variance. While the Euler contrast \eqref{PIeqn9} is known to have substantial bias when the time between observations is large, estimators such as \eqref{eqExact} are consistent for any value of $\delta$. The reason is that they form quadratic martingale estimating functions, see \cite{bibby1995martingale} and \cite{godambe1987quasi}.

An approximation for the variance of the maximum likelihood estimator $\hat{\Theta}=\argmax l_n(\Theta)$ is given by the observed information matrix
\begin{equation*}
I_o = \left. -\frac{\partial^2 l_n(\Theta)}{\partial \Theta^\top\partial\Theta} \right|_{\Theta=\hat{\Theta}}
\end{equation*}
which may be calculated by numerical differentiation of the log-likelihood function at estimated values. The approximative covariance matrix of $\hat{\Theta}$ is then given by the inverse
$\Sigma_{\hat{\Theta}} = I_o^{-1}$
and an estimated standard error of the $j^\text{th}$ parameter by $\sqrt{(I_o^{-1})_{jj}}$. 
Hence, an approximate $1-\alpha$ confidence region for $\Theta$ is given by the ellipse 
\begin{equation*}
\left\{\Theta: (\Theta-\hat{\Theta}) \Sigma^{-1}_{\hat{\Theta}}(\Theta-\hat{\Theta})^\top\leq\chi^2_d(1-\alpha) \right\}
\end{equation*}
where $d$ is the dimension of the row vector $\Theta$ and $\chi^2_d(1-\alpha)$ is the $1-\alpha$ quantile of the chi-square distribution with $d$ degrees of freedom.

{Finally we note that several additional approaches for statistical estimation of  Heston's model are known in the literature. For example, with access to high-frequency returns data, one may consult} \cite{barczy2016asymptotic} or \cite{de2015weighted}.

\subsection{Empirical study}
We base the empirical study on market data sourced from the Oxford-Man Institute of Quantitative Finance and from Wharton Research Data Services.\footnote{
\url{http://realized.oxford-man.ox.ac.uk} 
and \url{https://wrds-web.wharton.upenn.edu/wrds/}.} Data from Oxford-Man is used for the {price} of the S\&P 500 index, both historical closing prices observed with a daily frequency and high-frequency ($\sim5$min) returns which are used in the (pre-calculated) estimation of historical volatility. Data from the Option Metrics database, sourced through Wharton Research, is used for the price of European call options written on the S\&P 500 index. Here we use historical quotes of bid and offer prices from options with different strike-prices and maturities along with 
the relevant dividend yield paid by the underlying index and the risk-free interest rate that corresponds to the maturity of each option in the  set.

Prior to the numerical calculation of pricing bounds for the call options, we estimate the parameters of Heston's model from the S\&P 500 price and variance according to the following steps:
\begin{enumerate}
\item
First, we decimate the observation frequency of the variance to weekly observations by calculating the realized variance measure over week-long intervals, see Figure \ref{PIfig7}. This operation smooths the measured variance process and, in particular, it removes the extreme variance spikes (cf. Figure \ref{PIfig6}) which might cause non-robust parameter estimates.
\begin{figure}[t!]
\makebox[\textwidth][c]{  
\includegraphics[scale=0.44,trim=0 20 20 20,clip]{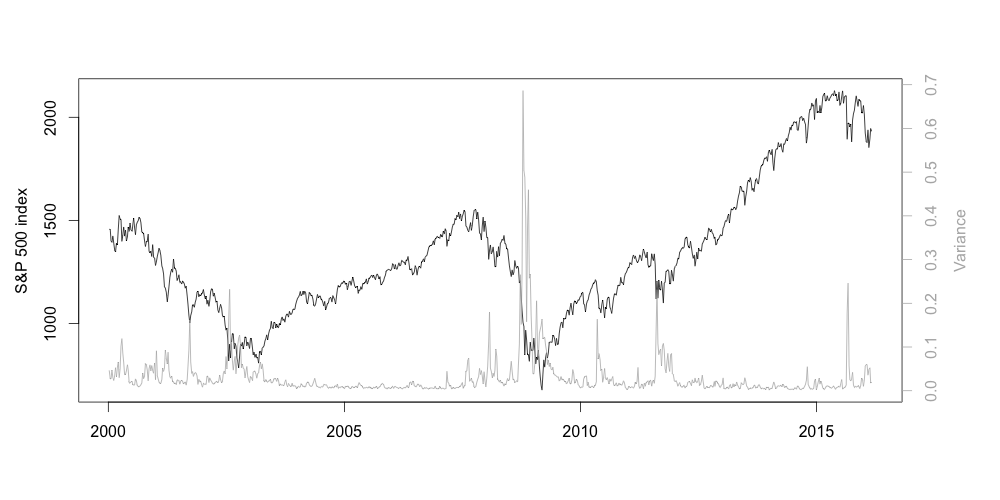}
}
\caption{Historical closing prices and realized variances of the S\&P 500 index, 843 weekly observations from January 3\textsuperscript{rd}, 2000 to February 29\textsuperscript{th}, 2016.}
\label{PIfig7}
\end{figure}
\item
We estimate $(\kappa,\beta,\sigma)$ from the weekly variance with the parametrisation $(\kappa,\beta,\sigma)\mapsto(\kappa,\theta,\sigma)$ of the model. We employ the approximative likelihood based on Euler conditional moments\footnote{Alternatively, we may employ the (approximative) likelihood with exact conditional moments. For daily observations, the numerical optimisation does not converge while for weekly data, this yields very similar parameter estimates and standard errors as with approximative moments.} and calculate the approximative covariance matrix accordingly by numerical differentiation. Results are given in Table \ref{PItab13} (with squared elements of the covariance matrix for a notion of standard errors) together with estimation results from the daily variance. From the results in Table \ref{PItab13} note that the daily variance yields relatively high estimates of the mean-reversion speed to accommodate extreme observations and also large standard errors of both drift parameters, which indicate that the square root process is a poorly fitting model for the daily variance data. With the weekly variance we obtain more sensible results and lower standard errors.
\item
In addition to estimated parameters of the variance process, we estimate the correlation coefficient of the model with a realised covariation measure.\footnote{The quadratic covariation of logarithmic data gives $\frac{1}{t}[\log S ,\frac{1}{\sigma}\log V ]_t = \frac{1}{t}\int_0^t\sqrt{V_s}\frac{1}{\sqrt{V_s}}d[\rho W^1+\sqrt{1-\rho^2}W^2,W^1]_s = \rho$ and we use a realized covariation estimate thereof.} This gives an estimate $\rho = -0.274$ from the weekly variance and closing price of the S\&P 500 index.
\end{enumerate}
\begin {table}[h!]
\begin{center}
\begin{tabular}{| c  c  c  c |} 
\multicolumn{4}{c}{Estimates, daily data} \\
\hline
& $\kappa$  & $\theta$ & $\sigma$  \\
\hline
& 29.6 & 0.0315 & 2.58 \\ 
 \hline 
\multicolumn{4}{c}{} \\
 \multicolumn{4}{c}{Standard errors} \\ 
 \hline 
 & $\kappa$  & $\beta$ & $\sigma$ \\ 
\hline
 $\kappa$ & 4.314 & 0.544 & 2.37e-06 \\ 
 $\beta$ & 0.544 & 0.074 & 1.79e-06 \\ 
 $\sigma$ & 2.37e-06 & 1.79e-06 & 0.0288 \\
\hline
\end{tabular}\hspace{15pt}
\begin{tabular}{| l  c  c  c |} 
\multicolumn{4}{c}{Estimates, weekly data} \\
\hline
& $\kappa$  & $\theta$ & $\sigma$  \\
\hline
& 4.59 & 0.0307  & 0.775 \\ 
 \hline 
\multicolumn{4}{c}{} \\
 \multicolumn{4}{c}{Standard errors} \\ 
 \hline
 & $\kappa$  & $\beta$ & $\sigma$ \\ 
\hline
 $\kappa$ & 1.395 & 0.621 & 1.91e-06 \\ 
 $\beta$ & 0.621 & 0.0269 & 4.84e-06 \\ 
 $\sigma$ & 1.91e-06 & 4.84e-06 & 0.0189 \\
\hline
\end{tabular}
\caption{Parameters and standard errors estimated from historical data of the S\&P 500 index. \textbf{Left table:} results based on 4,035 daily observations. \textbf{Right table:} results based on 843 weekly observations decimated from the original data. All estimates from numerical optimisation and differentiation of the approximative likelihood function based on Euler moments.}
\label{PItab13}
\end{center}
\end{table}

With estimated model parameters and standard errors in hand, we continue to the calculation of upper and lower pricing bounds for European options. We proceed according to the following:
\begin{enumerate}
\item
We consider call options on the S\&P 500 index with historical market prices from the three-year period of August 31\textsuperscript{st}, 2012 to August 31\textsuperscript{st}, 2015. 
We select the dates from this period that coincides with the weekly index data (i.e. dates for which both the option and the S\&P 500 closing price are quoted). This results in 157 dates 
and a total of 244,239 option quotes for different strikes and maturities.
\item
For each of the 157 dates during the time period, we chose a single option with strike price and time to maturity as shown in the right pane of Figure \ref{PIfig8}. Here, the ``initial" option of each period is selected with a medium-seized maturity and a strike-price as close as possible to being at-the-money. We then retain the same maturity and strike price (the same option) as far as there is  market quotes available. This gives us four options in total. We further record the relevant risk-free rate as given by the  zero-coupon interest rate with corresponding maturity, and the current dividend yield. The left pane of Figure \ref{PIfig8} shows the resulting bid/offer quotes after they have been converted to zero-dividend prices. This is done for each quote by first calculating the Black-Scholes implied volatility (with the effective dividend yield) and then recalculating the Black-Scholes price with zero dividend.
\begin{figure}[t!]
\makebox[\textwidth][c]{  
\includegraphics[scale=0.42,trim=0 0 20 20,clip]{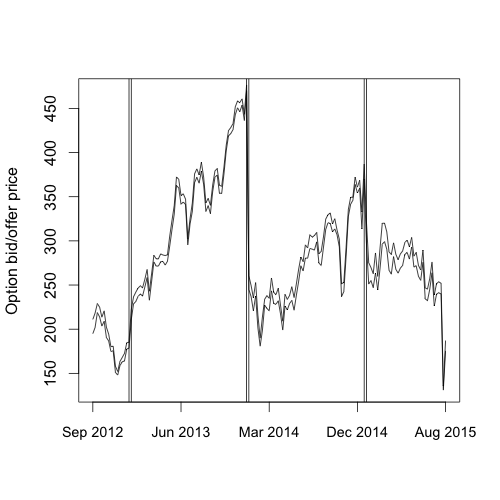}
\includegraphics[scale=0.42,trim=0 0 20 20,clip]{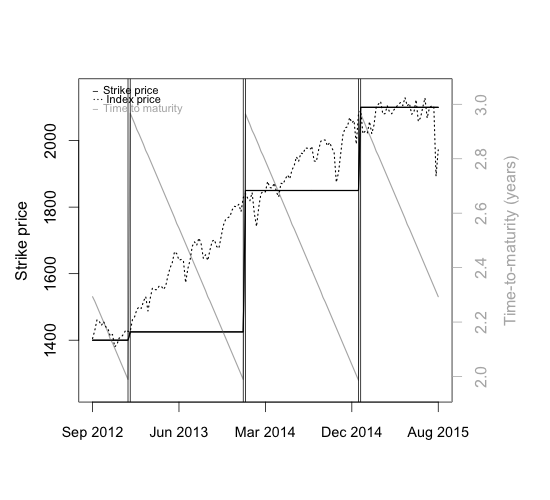}
}
\caption{\textbf{Left figure:} historical market prices of call options on the S\&P 500 index. The figure shows 157 bid/offer quotes (converted to zero-dividend prices) from the period August 31\textsuperscript{st}, 2012 to August 31\textsuperscript{st}, 2015. 
\textbf{Right figure:} the strike price and time-to-maturity of the call options.}
\label{PIfig8}
\end{figure}
\item
For the calculation of pricing bounds for the call option we employ a numerical method described in the Appendix. Here we initially have the following numerical considerations: Firstly, for the parameter estimates in Table \ref{PItab13} (based on weekly data) we have $\sqrt{4\beta}=0.751<0.775=\sigma$ which implies that the implicit Milstein scheme may fail due to the generation of negative outcomes (see Appendix \ref{AppHe}). To prevent this, we include a truncation step\footnote{The time-stepping of the scheme fails whenever $V^{\pi}_{t_i}<0$ due to the computation of $\sqrt{V^{\pi}_{t_i}}$ and the truncation step is simply to replace with $\sqrt{(V^{\pi}_{t_i})^+}$. Although this prevents the scheme to fail, note that negative values may still be generated and in particular when the time-step $\Delta_i$ is large. } to the simulation  according to the suggested method of \cite{andersen2010}. Further, we increase the number of time steps to $n=1,000$ for the forward simulation to prevent the generation of negative variance values.\footnote{Bookkeeping the sign of the generated variance values yields positive outcomes $99.3\%$ of the time when using $n=1,000$ time steps and $96.7\%$ for $n=25$.} We then down-sample the simulated price and variance to the original time-grid of $n=25$ steps for the backward simulation. Secondly, the driver of the backward process will explode for variance values approaching zero. As the forward simulation may output negative/zero values, we cancel the control: $u(X_t,Z_t,Y_t)\equiv0$ giving $H(X_t,Z_t,Y_t) = -rY_t$, each time the variance is smaller than a threshold, $V_t<\varepsilon$, and we set $\varepsilon = 0.00041$, which is the minimum value of the S\&P 500 variance.
\item
We simulate the optimally controlled value-process by the explicit scheme with $Y$-recursion and the MARS method of degree 2 for the regressions, see Appendix \ref{AppNS}. The control variate is included and we simulate $N=100,000$ paths of the forward-backward process over a time grid with $n=25$ time steps (with $n=1,000$ down-sampled to $n=25$ for the forward process). 
For each option price, we run a separate simulation with the estimated model parameters and corresponding maturity/strike, risk-free rate and initial values for the forward process (S\&P 500 index level and variance) from the market data. We simulate the backward process with the minimised driver (for the lower pricing bound) and the maximised driver (for the upper bound) based on the same forward simulation. Each evaluation of the optimal driver is calculated with the covariance matrix $\Sigma_{r,\kappa,\beta}$ where we use estimated standard errors and correlation for $\kappa,\beta$ from Table \ref{PItab13} and a standard deviation of 0.00005 for the interest rate ($r$ uncorrelated with $\kappa,\beta$). As before, we use a confidence level of $95\%$ for the uncertainty region.
\item
Finally, we also calculate the corresponding minimum/maximum prices for each considered call option by numerical optimisation of Heston's pricing function over the same $95\%$ uncertainty region---see Appendix \ref{AppHe} for details. Henceforth, we refer to these as formula-optimal prices.
\end{enumerate}

\begin{figure}[t!]
\makebox[\textwidth][c]{  
\includegraphics[scale=0.44,trim=0 20 20 20,clip]{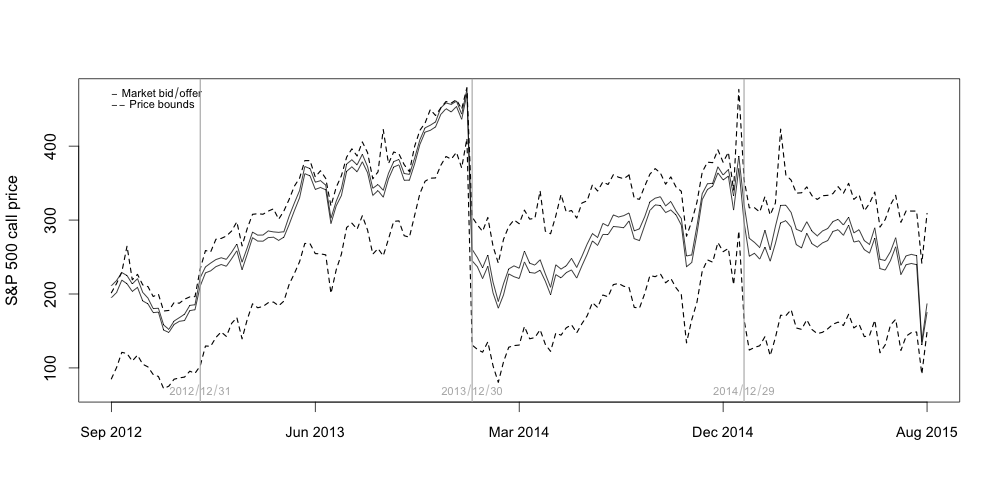}
}
\caption{European call options on the S\&P 500 index: upper and lower pricing bounds (dashed lines) as calculated by simulations of the optimally controlled value process. The graph shows model prices and historical market quotes of bid and offer prices (solid lines) on a weekly basis of options with four different strike/maturity structures (see right pane of Figure \ref{PIfig8}).  }
\label{PIfig9}
\end{figure}

\begin{figure}[t!]
\makebox[\textwidth][c]{  
\includegraphics[scale=0.44,trim=0 20 20 20,clip]{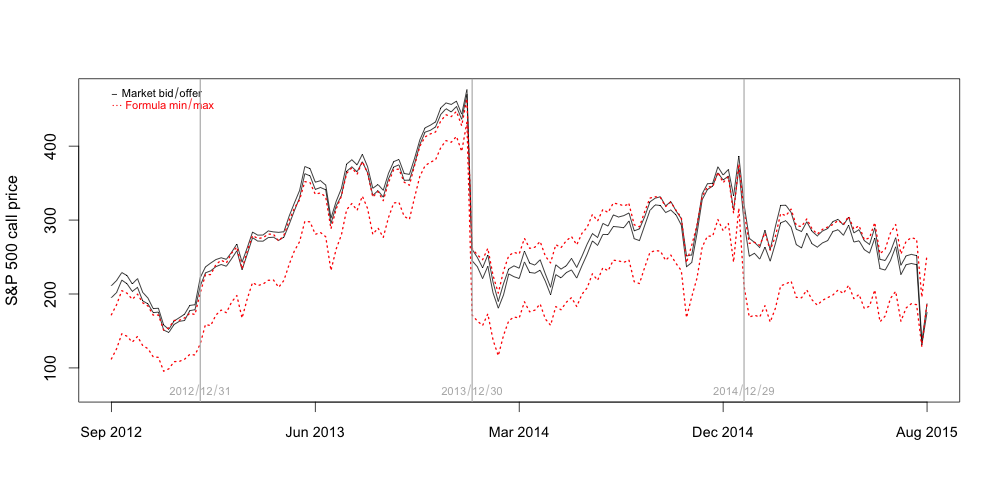}
}
\caption{The minimum and maximum price (red dotted lines) as obtained from the optimisation of Heston's pricing formula for call options. The solid black lines show the market bid/offer quotes of the S\&P 500 options. }
\label{PIfig10}
\end{figure}

The resulting upper and lower pricing bounds from the simulations are presented in Figure \ref{PIfig9} together with the corresponding market quotes of bid and offer prices. The formula-optimal prices are depicted in Figure \ref{PIfig10}. The dates at which there is a common change in strike price and maturity (see right pane of Figure \ref{PIfig8}) have been marked in the figures to separate the different options.
We label these options (I)-(IV) and give some additional results in Table \ref{PItab15}.
\begin {table}[h!]
\makebox[\textwidth][c]{
{\footnotesize
\begin{tabular}{| c |  c c c c |} 
\multicolumn{5}{c}{\textbf{European call options on the S\&P 500 index}} \\
\hline
\text{Option:} &  (I) & (II) & (III) & (IV)  \\
\hline
\hline
Duration 		& 12/09/04-12/12/24 & 12/12/31-13/12/23 &  13/12/30-14/12/22   &  14/12/29-15/08/31    \\
Maturity 		& 14/12/20 	& 15/12/19 & 16/12/16   &   17/12/15 \\
Spread:price	& $5.6\%$   	&  $3.4\%$	   & $5.6\%$	&  $7.1\%$  \\
\hline
Bounds:spread   & 14.9   		&  $12.4$ & 13.9	& 12.8		\\
In bounds 		& $88.2\%$   	&  $98.1\%$ & $100\%$ & $100\%$ \\
\hline
Optim:spread 	& 7.8			& 6.1			& 7.2 		& 6.6  \\
In interval 		& $11.8\%$  	& $0\%$		& $76.9\%$	& $72.2\%$  \\
\hline
\end{tabular}
}
}
\caption{Key figures for the model prices of S\&P 500 call options from the optimally controlled value process (the pricing bounds) and for the formula-optimal price interval. The spread-to-price ratio gives the average size of the market spread as a percentage of the average mid-market price of each option. Similarly, the bounds-to-spread ratio compares the range of the pricing bounds to that of the market spread.
The in-bounds figures give the proportion of market quotes that fall inside the model bounds. Corresponding figures are calculated for the intervals of the formula-optimal prices. }
\label{PItab15}
\end{table}

\begin{figure}[t!]
\centering
\includegraphics[scale=0.43,trim=0 0 20 20,clip]{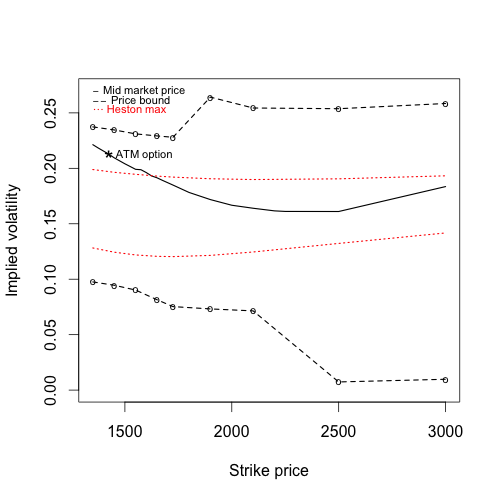}
\includegraphics[scale=0.43,trim=0 0 20 20,clip]{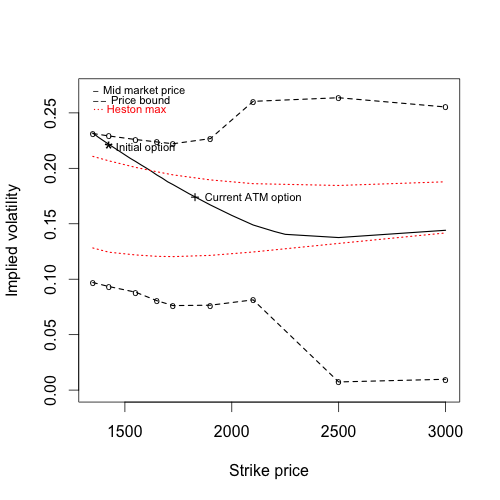}
\caption{\textbf{Left figure:} mid-market implied volatility of the S\&P 500 call option for different strikes as recorded on 2012-12-31. Corresponding model-boundaries (dashed lines) and formula-optimal prices (red dotted lines), both in terms of implied volatilities. The volatility of the ATM option (II) is marked with a star. \textbf{Right figure:}  implied volatilities from the mid-market price, model-boundaries and formula-optimum, as recorded on 2013-12-23 (the last date of the considered period for option (II)).
}
\label{PIfig11}
\end{figure}

\noindent A note on the interpretation of these results are in order here.
\begin{itemize}
\item
First, we see that the market bid/offer quotes fall inside the model bounds for almost all considered call prices (154 out of 157) and in particular, for all prices when looking at the latest two options (III)-(IV), see Table \ref{PItab15}. The lower bound is always covering the bid quote and the price interval of the bounds is fairly symmetrical around the mid-market price for options (III)-(IV). 

\item
For option (I), the offer prices are close to the upper bound (occasionally above) and the same holds for option (II). This option's moneyness is strongly increasing with time (see right pane of Figure \ref{PIfig8}) while the distance from the upper bound to the offer quote is shrinking. A possible explanation may be that the model is unable to capture the slope and skew of market prices/implied volatilities for the parameters we use, in particular since these are estimated from historical index prices and not from option prices.
\end{itemize}
For an investigation of this point, we take a look at the market prices and model boundaries for a range of strikes at the first and last date of option (II), plotted in terms of implied volatilities in Figure \ref{PIfig11}. 

\begin{itemize}
\item
The left pane of Figure \ref{PIfig11}, which shows prices form the starting date, shows a strongly skewed volatility curve for the market prices, while the prices from Heston's optimised formula yield much flatter, less sloped, curves. This indicates that  a higher level of skewness is required to fit the curvature of market volatilities (roughly speaking, a stronger negative correlation to increase the slope and a higher level of ``vol-of-vol'' to increase the skew), and a higher volatility level overall for the ATM formula-price to fit with the market (a higher mean-reversion level and lower reversion speed).

\item
On the other hand, the corresponding pricing-boundaries are wide enough to cover all strikes at both dates, even if the upper boundary is close to the market volatility at the later date where the option is deeply in-the-money---see right pane of Figure \ref{PIfig11}. 
As in the case with formula-optimal prices, we note that the bounds do not exhibit any curvature in line with the market volatilities, supposedly because of the low level of negative correlation and vol-of-vol. 
\end{itemize}

Returning to the prices of all options (I)-(IV), we have in total that 98\% of the market quotes are within the pricing bounds whilst the  formula optimal prices cover $40\%$ of the quotes. In particular, the market quotes of option (II) are outside the formula-optimal prediction throughout the period, and option (I) has only $11.8\%$ of its quotes covered. Generally, the upper price is too low and the coverage of the formula-price is to small:  the (average) ranges of the optimised prices are $\sim7$ times the sizes of market spreads (see Table \ref{PItab15}), and thus almost halved compared to the ranges of the model bounds ($\sim13$ times the market spreads).

The optimisation of Heston's formula based on statistical inference is not sufficient to cover for the market quotes of the considered data: the price-interval is too tight, and options in-the-money generally fall outside the model predictions, which indicated that the volatility smile is not captured. This should perhaps be expected since parameters are estimated to fit the underlying index---diffusion parameters in particular---and not to fit the actual options we are trying to price. Further, we use a constant set of parameters to predict option prices over the whole three-year period, while in practice one would typically update the estimates on regular basis. {We have thus faced Heston's model with a challenging task: to price a dynamical set of market options over a long time period while taking in information from the underlying asset alone when estimating the model to data.} In return, we allow drift parameters to vary within their 95\% confidence region as a representation of the incompleteness of the market model which, after all, gives an optimised price range that cover option quotes to some extend. Only when we generalise the model 
we obtain conservative pricing bounds wide enough to cover most prices, even if some deep in-the-money options still fall outside. We assume the same 95\% confidence region as a representative for the incompleteness, {but allow for a much more general view on the uncertainty of the parameters which span the space of risk-neutral measures: not only are they uncertain within the confidence region, the parameters also change dynamically with time. In the end, the former method corresponds to an optimally controlled value process with parameter processes being constants, and we simply have to allow for these parameters to vary in order to cover the market pricing of options in a satisfactory way.}

\section{Stochastic volatility models with jumps}\label{PIsecG}
For the purpose of completion, we will generalise our modelling framework in this final section to a multi-asset setting under a  Markovian stochastic volatility model with jumps. Our intension is to give a brief presentation of how the uncertainty pricing transfers to a general model and we deliberately avoid going too deep into details and technical assumptions.

\subsection{A generic Markovian model}

We consider a financial market model on a filtered probability space $(\Omega,\mathcal{F},\{\mathcal{F}_t\}_{t\geq0},P)$ that consists of a money account $B$, paying a risk-free interest of deterministic rate $r$, and a $\R^{d}$-valued stochastic process $S=(S^1,\dots,S^d)^\top$ representing the price processes of $d$ risky assets.  Furthermore, we have $d'$ non-negative stochastic processes, $V$ taking values in $\R^{d'}$, that represents the instantaneous variances. Typically, the two are of equal dimension such that each asset price is diffused by an individual volatility and we also assume that $d'=d$ is the case here. The statistical $P$-dynamics of the $m=2d$ column-vector $X=(S;V)$ of state variables are assumed to be of the form
\begin{equation*}
\begin{aligned}
dX_t = \mu^p(X_t)dt + \sigma(X_t)dW_t + \int_{\mathcal{Z}}h(\xi,X_{t-})\tilde{\mu}(d\xi,dt)
\end{aligned}
\end{equation*}
where $\mu^p(\cdot)$ is the $m$-dimensional drift-function under the statistical measure, $\sigma(\cdot)$ the $m\times m$-valued diffusion matrix, and $W$ is a $\R^m$-valued Wiener process. The jump part is driven by $\tilde{\mu}$, a compensated Poisson random measure on a Blackwell space $\mathcal{Z}$ with deterministic compensator $\mu_p(d\xi,dt)=\nu(d\xi)dt$, and $h(\cdot)$ is a state-dependent function valued in $\R^m$ that {governs the jump sizes of $X$}. Since we are working with a $\R^m$-dimensional state processes, we take $\mathcal{Z}=\R^m$. We assume that $\{\mathcal{F}_t\}_{t\geq0}$ is generated by $W$ and $\tilde{\mu}$ jointly, and augmented to satisfy the usual conditions. The functions $\mu^p$, $\sigma$, $h$ are assumed to be defined such that the SDE admits a {unique} solution {up to a fixed deterministic time $T$} (for instance {of linear growth and locally Lipschitz continuous}), and $V$ being non-negative almost surely. Further, we assume sufficient integrability conditions such that the market model admits no arbitrage: there exists an equivalent martingale measure $Q$ under which $S$ and $V$ follows 
\begin{equation*}
\begin{aligned}
&dS_t = rS_tdt + \sigma_S(S_t,V_t)d\tilde{W}_t +  \int_{\R^d}h(\xi,S_{t-},V_{t})\tilde{\mu}(d\xi,dt) \\
&dV_t = \mu_V(V_t,\Gamma)dt + \sigma_V(V_t)d\tilde{W}_t
\end{aligned}
\end{equation*}
where $\sigma_S(\cdot)$ and $\sigma_V(\cdot)$, both with values in $\R^{d\times m}$, are the first and last $d$ rows of $\sigma$. The $\R^d$-valued function $\mu_V(\cdot,\Gamma)$ is the $Q$-drift of the variance with parameters $\Gamma$, and $\tilde{W}$ is a $m$-dimensional Wiener process under $Q$. For our convenience, we have assumed that jumps affect the asset prices\footnote{As the jumps are generated by a Poisson random measure, $S$ will have jumps given by 
\[
\Delta S_t=\int_{z\in\R^d} h(z,S_{t-},V_t)\tilde{\mu}(dz,\{t\}) = h(z_t,S_{t-},V_t)\mathbf{1}_{\{\Delta S_t\neq 0  \}}
\]
where $z_t\in\R^d$ is a (unique) point in the set where $\mu(\{z_t\})=1$. } only; $\mathcal{Z}=\R^d$ and $h(\cdot)$ is an $\R^d$-valued function while the compensator measure  
$\mu_p(d\xi,dt)=\nu(\gamma,d\xi)dt$ is dependent on $Q$-parameters $\gamma$ (we use the same notation for $\mu_p$ here even if the compensator may be different under $P$ and $Q$). Furthermore, we assume that the continuous variance has drift and diffusion functions dependent on the state of the variance alone. The risky assets are not assumed to carry any dividend payments, although the generalisation to non-zero dividends (as well as jumps and $S$-dependent coefficients for the variance) should be straightforward. As under $P$, we assume all coefficients under $Q$ to sufficiently well behaved for an appropriate solution to exists.

The market model $(S,B)$ is free of arbitrage but incomplete as it has more random sources ($2\times d$) than traded risky assets ($d$), {and since the asset prices exhibit jumps} (i.e. the risk-neutral measure $Q$ is not unique). For a Markovian pricing rule 
\begin{equation*}
\begin{aligned}
D_t &= D(t,S_t,V_t),\,\,\,\,\,t\in[0,T], \\
D &: [0,T]\times\R^d\times\R^d \rightarrow \R,\,\,\,\,\,D\in\mathcal{C}^{1,2},
\end{aligned}
\end{equation*}
of a European option with terminal payoff $g(S_T)$, we have a pricing equation corresponding to (\ref{PIeqn31}) as given by 
\begin{equation*}
\begin{aligned}
\frac{\partial D}{\partial t} + \mathcal{L}D -rD &= 0 \\
D(T,s,v) &= g(s)
\end{aligned}
\end{equation*}
where $\mathcal{L}$ is the (time independent) inegro-differential operator that generates $(S;V)$ under $Q$. For a function $f(s,v)\in\mathcal{C}^2$ the operator is defined as
\begin{equation*}
\begin{aligned}
\mathcal{L}f(s,v) &= \mu^Q(s,v)^{\top}\nabla_xf(s,v) + \frac{1}{2}\text{tr}\left[  \sigma^{\top}(s,v)\nabla^2_{xx}f(s,v)\sigma(s,v)  \right] \\
&+ \int_{\xi\in\R^d}\left( f(s+h(\xi,s,v),v)-f(s,v) - h(\xi,s,v)^{\top}\nabla_xf(s,v)   \right)\nu(d\xi)
\end{aligned}
\end{equation*}
where $\nabla_x$, $\nabla^2_{xx}$ with $x=(s,v)$ are the gradient and Hessian operators respectively, 
and $\text{tr}[\cdot]$ the matrix trace. By the Feynman-Kac representation formula, this is equivalent to the risk-neutral valuation formula
$D(t,s,v) = \E^Q\left[\left. e^{-r(T-t)}g(S_T)\right| (S_t,V_t)=(s,v) \right]$.

\subsection{Pricing under parameter uncertainty}
Here we introduce  the controlled measure $Q^u$ that represents the parameter uncertainty in our model. Hence, with $\{u_t\}_{t\geq0}$ being a $\mathcal{F}_t$-predictable control process that takes its values in a compact uncertainty set $U\subset \R^k$, we let the controlled dynamics be
\begin{equation*}
\begin{aligned}
&dS_t = r_tS_tdt + \sigma_S(S_t,V_t)dW^u_t +  \int_{\R^d}h(\xi,S_{t-},V_{t})\tilde{\mu}^u(d\xi,dt) \\
&dV_t = \mu_V(V_t,u_t)dt + \sigma_V(V_t)d{W}^u_t.
\end{aligned}
\end{equation*}
where we assume that the controlled drift of the asset price and variance, $r_tS_t$ and $\mu_V(V_t,u_t)=\mu_V(V_t,\Gamma_t)$, to be of the same functional form under $Q$ and $Q^u$. Hence, the control has components $u_t = (r_t,\Gamma_t,\gamma_t)$ where $\Gamma$ are the parameters of $\mu_V$, 
while $\gamma$ are parameters to the controlled (form invariant) compensator measure
\[
\mu_p^u(d\xi,dt)=\nu(\gamma_t,d\xi)dt
\] 
with Radon--Nikodym density $\beta(\xi,t,u_t)\equiv d\mu^u_p/d\mu_p$ with respect to $\mu_p$.
We let $\mu^{Q^u}(S_t,V_t,u_t) \equiv (r_tS_t;\mu_V(V_t,u_t))$ denote the common drift of $(S;V)$ under $Q^u$ (and similarly for the common $Q$-drift). The effect of the control is then defined by the $\R^{1\times m}$-valued process 
{\footnotesize
\begin{equation*}
\alpha(S_t,V_t,u_t) = \sigma^{-1}(S_t,V_t)\left(\mu^{Q^u}(S_t,V_t,u_t)-\mu^Q(S_t,V_t) -\int_{\R^d}h(\xi,S_{t-},V_t)(\beta(\xi,t,u_t)-1)\nu(d\xi)
\right)
\end{equation*}
} 
to give the linear driver function $f(s,v,y,z,\theta,u) = -ry + z\alpha(s,v,u)+\int_{\xi}\theta(\xi)(\beta(\xi,t,u)-1)\nu(d\xi)$ where the second last argument is a function $\theta:\R^d\mapsto\R$. 
Modulo sufficient integrability and Lipschitz conditions, we have that the value function for a fixed admissible control $J_t(u)=\E_u[e^{-\int_t^Tr_udu}g(S_T)|\mathcal{F}_t]$, $t\in[0,T]$, is given as part of the solution $(J(u),Z,\Theta)$ to the linear BSDE
\begin{equation*}
\begin{aligned}
dJ_t(u) &= -f(S_t,V_t,J_t(u),Z_t,\Theta_t,u_t)dt+Z_td\tilde{W}_t + \int_{\xi\in\R^d}\Theta_t(\xi)\tilde{\mu}(d\xi,dt) \\
J_T(u) &= g(S_T)
\end{aligned}
\end{equation*}
where $Z$ is $\R^{1\times m}$-valued while $\Theta$ is a process taking its values in the space of functions $\theta:\R^d\mapsto\R$. 
The result follows similarly as in the case of Heston's model: apply It\^{o}'s product rule to $\mathcal{E}(\Lambda)J(u)$, where $\Lambda = -\int_0^.r_tdt + \alpha(S_t,V_t,u_t)\bullet \tilde{W} + \int_0^.\int(\beta(\xi,t,u_t)-1)\tilde{\mu}(d\xi,dt)$, to see that $\mathcal{E}(\Lambda)J(u)$ is a martingale, and use that  $\mathcal{E}(\Lambda+\int_0^.r_tdt)_T=dQ^u/dQ$ for  the measure change of $\mathcal{E}(\Lambda)_tJ_t(u) = \E[\mathcal{E}(\Lambda)_TJ_T(u)|\mathcal{F}_t]$ to obtain the original expression for the value process after rearrangement.
Further, defining the pointwise optimised driver functions over the compact uncertainty set
\begin{equation*}
H^{\pm}(s,v,y,z,\theta) = \left|  \esssup_{u\in U} \pm f(s,v,y,z,\theta,u)    \right|,
\end{equation*}
we have by the comparison theorem that 
the the optimally controlled value processes (the upper/lower pricing boundaries) $\{D^{\pm}_t\}_{t\in[0,T]}=\{  |\esssup_{\{u_t\}} \pm J_t(u)| \}_{t\in[0,T]}$ are solutions to the BSDEs
\begin{equation*}
\begin{aligned}
dD^{\pm}_t &= -H^{\pm}(S_t,V_t,D_t^{\pm},Z_t,\Theta_t)dt + Z_t d\tilde{W}_t + \int_{\xi\in\R^d}\Theta_t(\xi)\tilde{\mu}(d\xi,dt) \\
D^{\pm}_T &= g(S_T).
\end{aligned}
\end{equation*}
Here as well, this is a consequence of the fact that we have a linear driver in $y$, $z$ and $\theta$, and from the comparison theorem for BSDEs. The proof, which we omit for brevity, follows in the same fashion as in the previous case with Heston's model (see \cite{cohen2015stochastic}, Chapter 21, for details). As well, since we work in a Markovian setting, we have that the solution can be written with a deterministic function $D_t = D(t,S_t,V_t)$, and the same holds for the optimal control: there exists a function $u^*(t,s,v)$ such that the feedback control $u^*_t=u^*(t,S_t,V_t)$ is the optimal control among all admissible controls.

Finally, as in the case with Heston's model, we have by the semilinear Feynman--Kac formula that $D(t,s,v)$  satisfies a semilinear partial differential equation
{\footnotesize
\vspace{10pt}
\begin{equation*}
\frac{\partial D}{\partial t} + \frac{1}{2}\text{tr}[\sigma^\top\nabla^2_{xx}D\sigma] 
+ \essinf_{(r,\Gamma,\gamma)\in U}\left\{ -rD+(\mu^{Q^u})^{\top}\nabla_xD + \int_{\xi} ( \Delta D - h^{\top}\nabla_xD)\nu(\gamma,d\xi)   \right\} = 0
\vspace{10pt}
\end{equation*}
}with terminal condition $D(T,s,v)=g(s)$, and where $\Delta D$ is shorthand notation for $D(t,s+h(\xi,s,v),v)-D(t,s,v)$. Although many numerical methods exist for a PIDE of this type, one may opt for simulating the BSDE solution instead (see e.g. \cite{bouchard2008discrete}), especially when the dimensional of the problem is high.

\section{Conclusion} \label{PIsecC}
Model uncertainty, here represented by parameter uncertainty, is an acknowledged concept formalised by \cite{knight1921risk} and its importance has been studied in the financial context at least since \cite{dermanMR}. 
The focus of this paper has been to investigate how parameter uncertainty could be incorporated into a stochastic volatility model, and how it affects derived prices of European option. The considered uncertainty was fairly general: interest rate and volatility drift parameters where allowed to change over time (constant-parameters being a special case) 
within a pre-described uncertainty region inferred from statistical estimation. The effect on pricing was then studied from a worst-case perspective with boundaries for the option price that could be embedded into a control problem, with the control playing a role of the uncertain parameters. 

With Heston's model as a working example, the control problem--BSDE duality was then exploited and an explicit equation for the pricing boundary (the optimal value process) was derived in the form of a Markovian linear BSDE. A numerical scheme with several suggested modifications was considered for the solution of this BSDE, and an evaluation of the schemes was made in a known-outcome setting analogous to the dynamic-parameter setting. Based on bias/variance (and computational) considerations, a scheme was proposed for an empirical study of the methodology applied to real-world market data. Studying a set of bid/offer market quotes of European call options on the S\&P 500 index and their corresponding model-price bounds, it was found that even if the model (and uncertainty set) was estimated from historical prices of the underlying, 98\% of the market option prices was within the model-prescribed bounds. In contrast, $\sim40\%$ of the market quotes was within the maximum/minimum model-price interval when constant parameters where used.

In both the dynamic and constant parameter setting, it was seen that the model implied volatilities did not follow the curvature of the market implied volatilities. A natural explanation for this observation is that the diffusion parameters, which effectively decide the slope and skew of the implied volatility curve, were estimated from asset-price data, and not from option-price data. An interesting empirical sequel would therefore be to study how the shape and coverage of model-price bounds change when parameters are calibrated from market option prices instead. We leave this for further investigation.

Finally, we note that prior beliefs and preferences about the uncertainty are not taken into consideration by the conservative approach with pricing boundaries. However, with $L(u_t,u'_t)$ being some function that assigns a loss when $u_t$ is used instead of the true parameters $u_t'$, we could incorporate beliefs with a value function of the form
\[
J_t(u) = \E_u\left[ \left. e^{\int_t^Tr_sds}G+\int_t^TL(u_s,u'_s)ds \right| \mathcal{F}_t \right]
\]
that would lead to a similar linear BSDE. The loss could be based on the (approximative) normality of estimated parameters or some quantity related to an economic value, for instance a hedging error. In both cases, the value of the loss must be related to the value of the option payoff, an intricate task that we leave for further research along with this approach.

\subsection*{Acknowledgements}
We thank two anonymous referees for detailed and helpful recommendations on this paper. Tegn\'{e}r gratefully acknowledges support from the Wallenberg Foundations and the Oxford-Man Institute for Quantitative Finance. Cohen gratefully acknowledges support from Oxford-Nie Financial Big Data Laboratory and the Oxford-Man Institute for Quantitative Finance.

\clearpage

\begin{appendix}

\section{Numerical methods for BSDEs}\label{PIseqNum}

The optimally controlled value process (or the value process for a fixed feedback control, i.e. $u_t=u(t,S_t,V_t)$ for a deterministic function $u$) is given by the solution to the decoupled forward-backward stochastic differential equation (\ref{PIeqn13})-(\ref{PIeqn2}). In general, there is not much hope of finding closed-form solutions to neither forward nor backward SDEs and one typically has to consider numerical methods. For our purposes, we consider the simulation technique by \cite{bouchard2004discrete}. 

\subsection{The simulation scheme by Bouchard and Touzi}

For a time gird $\pi:0=t_0<\cdots<t_n=T$, \cite{bouchard2004discrete} propose a method to generate a discrete-time approximation $(X^\pi,Y^\pi)$  of the solution to a decoupled equation with forward component $X$ and backward component $Y$. In the first part of the scheme, the forward  component $X^\pi$ is simulated over the time grid $\pi$ with a standard Euler-Maruyama approximation to generate $N$ paths of $X^\pi$ (see e.g. \cite{kloeden1992numerical}). The component $Y^\pi$ is then generated by the backward induction
\begin{equation}\label{PIeqn14}
\begin{aligned}
&Y^\pi_{t_n} = g(X^\pi_{t_{n}}) \\
&Z^\pi_{t_{i-1}} = \frac{1}{\Delta_i} \E\left[\left. Y^\pi_{t_i}\Delta W_{t_i}\right| X^{\pi}_{t_{i-1}}  \right] \\
&Y^\pi_{t_{i-1}} = \E\left[\left. Y^\pi_{t_i}+f(X^{\pi}_{t_{i-1}},Y^{\pi}_{t_{i-1}},Z^{\pi}_{t_{i-1}})\Delta_i\right| X^{\pi}_{t_{i-1}}  \right] 
\end{aligned}
\end{equation}
where $\Delta_i \equiv t_i-t_{i-1}$ and $\Delta W_{t_i} \equiv W_{t_i}-W_{t_{i-1}}$ are the $i$\textsuperscript{th} time- and Wiener increments from the generation of $X^\pi$. The last equation in (\ref{PIeqn14}) is obtained by applying $\E[\cdot|\mathcal{F}_{t_{i-1}}]$ to the following simple discretization of the BSDE
\begin{equation}\label{PIeqn14aa}
Y^\pi_{t_{i}}-Y^\pi_{t_{i-1}}=-f(X^{\pi}_{t_{i-1}},Y^{\pi}_{t_{i-1}},Z^{\pi}_{t_{i-1}})\Delta_i + Z^{\pi}_{t_{i-1}}\Delta W_{t_i}
\end{equation} 
and using the Markov property of $X$ and the fact that $Y_t$ and $Z_t$ are both deterministic functions of $X_t$ for all $t\in[0,T]$. The second equation for $Z$ is obtained similarly by multiplying (\ref{PIeqn14aa}) with $\Delta W_{t_i}$ and taking conditional expectations. 

For the backward induction (\ref{PIeqn14}) one has to compute the conditional expectations and to make the scheme operational, this is made with an approximation $\hat{E}[\cdot|X^{\pi}_{t_{i-1}}]$ of the regression function $\E[\cdot|X^{\pi}_{t_{i-1}}]$ based on simulated training data. That is, the data 
\[
\{Y^{\pi(j)}_{t_i},\Delta W^{(j)}_{t_i},X^{\pi(j)}_{t_{i-1}}\}_{1\leq j\leq N} 
\]
is used for the first regression in (\ref{PIeqn14}) where  
$X^{\pi(j)}$ is the $j$\textsuperscript{th} simulated path of $X^\pi$ and $Y^{\pi(j)}$ is the corresponding value from the induction of the previous time step. For the second regression, $\{Y^{\pi(j)}_{t_i},Z^{(j)}_{t_{i-1}},X^{\pi(j)}_{t_{i-1}}\}_{1\leq j\leq N}$ is used accordingly. 

As an example of a non-parametric regression estimator, it is suggested to use the Nadaraya-Watson weighted average for a kernel estimator. We conveniently employ the $k$-nearest neighbour kernel for this purpose: for $X^{\pi}_{t_{i-1}}$ and a generic $\xi\in\mathcal{F}_{t_i}$, each with simulated outcomes $\{X^{\pi(j)}_{t_{i-1}},\xi^{(j)}\}_{1\leq j\leq N}$, we approximate $\E[\xi|X^{\pi}_{t_{i-1}}=X^{\pi(j)}_{t_{i-1}}]$, $j=1,\dots,N$, with
\begin{equation} \label{PIeqn15}
\hat{E}\left[\xi \left| X^{\pi(j)}_{t_{i-1}}\right.\right] = \frac{\sum_{l=1}^N \xi^{(l)}\mathbf{1}\left(||X^{\pi(l)}_{t_{i-1}}-X^{\pi(j)}_{t_{i-1}}||\leq d_k^{(j)} \right)}{k+1}
\end{equation}
where $d_k^{(j)}$ is the distance between $X^{\pi(j)}_{t_{i-1}}$ and its $k$\textsuperscript{th} nearest neighbour, and $\mathbf{1}(\cdot)$ is the indicator function.  The regression (\ref{PIeqn15}) together with (\ref{PIeqn14}) yields an implicit simulation method and as a last step of the scheme, it is suggested 
to truncate $Z^\pi_{t_{i-1}}$ and $Y^\pi_{t_{i-1}}$ if one has appropriate (possibly $t$ and $X^{\pi}_{t_{i-1}}$ dependent) bounds for $\E[ Y^\pi_{t_i}\Delta W_{t_i}| X^{\pi}_{t_{i-1}}  ]$, $\E[ Y^\pi_{t_i}| X^{\pi}_{t_{i-1}}  ]$ and $Y_{t_{i-1}}$.

The $k$-nearest neighbours estimator (\ref{PIeqn15}) approximates the regression function in a local neighbourhood  with a constant. As such, it has low bias but high variance, 
and it suffers from the curse of dimensionality at the boundaries. For an alternative regression estimator, we consider the MARS method\footnote{We are using the R package ``earth" by \cite{earthpackage}.} (multivariate adaptive regression spines) which uses piecewise linear basis functions in an adaptive manner to approximate the regression function. The model has the linear form
\begin{equation}\label{PIeqn23}
\hat{E}\left[\xi \left| X^{\pi}_{t_{i-1}}\right.\right] = \beta_0 + \sum_{m=1}^M \beta_m h_m(X^{\pi}_{t_{i-1}})
\end{equation}
where each basis function $h_m(X)$ is constructed from a collection of paired
 piecewise-linear splines
\begin{equation*}
\mathcal{C} = \left\{(X^k-\eta)^+, (\eta-X^k)^+ :\eta\in\{X^{k,\pi(j)}_{t_{i-1}}\}_{j=1}^N,k=1,2 \right\}.
\end{equation*}
The \textit{knots}, $\eta$, are hence placed at any value in the set of $X$-observations (with superscript $k$ referring to the components of $X$). A basis function $h_m(X)$ is also allowed to be a $d$-times product of the functions in $\mathcal{C}$, where $d$ denotes a chosen model order. 

The model (\ref{PIeqn23}) is built up 
 sequentially by multiplying a current $h_m$ with a new function from $\mathcal{C}$ to form candidates: all pairs in $\mathcal{C}$ are tested, and only the term which yields the largest reduction of residual error is added to \eqref{PIeqn23}. Here, all coefficients $\beta_0,\beta_1,\dots$ are estimated in each step by least squares. The model building then continues until there is a prescribed maximum  number of terms $M$. Finally, the model is ``pruned" with a deleting procedure (again based on the \textit{smallest} increase of error) where the optimal number of terms is estimated by cross-validation (for details, see \cite{hastie2005elements}). 

\subsection{Modified simulation schemes}
As a first modification of the Bouchard-Touzi method, we consider an explicit version of the implicit scheme by replacing the second regression in  (\ref{PIeqn14}):
\begin{equation}\label{PIeqn16}
\begin{aligned}
&Y^\pi_{t_n} = g(X^\pi_{t_{n}}) \\
&Z^\pi_{t_{i-1}} = \frac{1}{\Delta_i} \E\left[\left. Y^\pi_{t_i}\Delta W_{t_i}\right| X^{\pi}_{t_{i-1}}  \right] \\
&Y^\pi_{t_{i-1}} = \E\left[\left. Y^\pi_{t_i}+f(X^{\pi}_{t_{i-1}},Y^{\pi}_{t_{i}},Z^{\pi}_{t_{i-1}})\Delta_i\right| X^{\pi}_{t_{i-1}}  \right]. 
\end{aligned}
\end{equation}
This comes from a discretization of the BSDE at the right time-point $Y_{t_i}$ instead of $Y_{t_{i-1}}$ and since $Y$ is a continuous process, the effect of using the value at the right time-point is vanishing as the time-grid becomes tighter. {The discretization is used by} \cite{Gobet06numericalsimulation}, 
 and the benefit is that this allows for an explicit calculation of $Y^\pi_{t_{i-1}}$ in the second regression step of each iteration. 

As an additional step, to obtain an implicit method with a fixed point procedure, we may employ (\ref{PIeqn16}) to get a first candidate $\tilde{Y}^\pi_{t_{i-1}}$ and supplement each step in the backward induction with a small number of implicit iterations of
\begin{equation}\label{PIeqn28}
\tilde{Y}^\pi_{t_{i-1}} = \E\left[\left. Y^\pi_{t_i}+f(X^{\pi}_{t_{i-1}},\tilde{Y}^{\pi}_{t_{i-1}},Z^{\pi}_{t_{i-1}})\Delta_i\right| X^{\pi}_{t_{i-1}}  \right],
\end{equation}
and keeping ${Y}^\pi_{t_{i-1}}=\tilde{Y}^\pi_{t_{i-1}}$ as our {final value for the next backward step ---see} \cite{Gobet2005}.

Secondly, {to improve the stability of the scheme}, we consider a modification of (\ref{PIeqn16}) based on the following recursion for the backward component 
\begin{equation*}
\begin{aligned}
Y^\pi_{t_{i-1}} &= Y^\pi_{t_i}+f(X^{\pi}_{t_{i-1}},Y^{\pi}_{t_{i}},Z^{\pi}_{t_{i-1}})\Delta_i - Z^{\pi}_{t_{i-1}}\Delta W_{t_i} \\
& = Y^\pi_{t_{i+1}} 
+ f(X^{\pi}_{t_{i-1}},Y^{\pi}_{t_{i}},Z^{\pi}_{t_{i-1}})\Delta_i 
+ f(X^{\pi}_{t_{i}},Y^{\pi}_{t_{i+1}},Z^{\pi}_{t_{i}})\Delta_{i+1} 
- Z^{\pi}_{t_{i-1}}\Delta W_{t_i} - Z^{\pi}_{t_{i}}\Delta W_{t_{i+1}} \\
& = Y^\pi_{t_{n}} + \sum_{k=i}^n f(X^{\pi}_{t_{k-1}},Y^{\pi}_{t_{k}},Z^{\pi}_{t_{k-1}})\Delta_k - Z^{\pi}_{t_{k-1}}\Delta W_{t_k}
\end{aligned}
\end{equation*}
such that we may write the explicit backward induction (\ref{PIeqn16}) as
\begin{equation}\label{PIeqn17}
\begin{aligned}
&Y^\pi_{t_n} = g(X^\pi_{t_{n}}) \\
&Z^\pi_{t_{i-1}} = \frac{1}{\Delta_i} \E\left[\left. \left(Y^\pi_{t_{n}} + \sum_{k=i+1}^n f(X^{\pi}_{t_{k-1}},Y^{\pi}_{t_{k}},Z^{\pi}_{t_{k-1}})\Delta_k\right)\Delta W_{t_i}\right| X^{\pi}_{t_{i-1}}  \right] \\
&Y^\pi_{t_{i-1}} = \E\left[\left. Y^\pi_{t_{n}} + \sum_{k=i}^n f(X^{\pi}_{t_{k-1}},Y^{\pi}_{t_{k}},Z^{\pi}_{t_{k-1}})\Delta_k\right| X^{\pi}_{t_{i-1}}  \right].
\end{aligned}
\end{equation}
{This idea appears in} \cite{bender2007forward} and is explored further in \cite{gobet2016linear}. The benefit 
is that {errors due to approximating the conditional expectation do not accumulate at the same rate}. As in the previous modification, we may complement (\ref{PIeqn17}) with a small number of iterations
\begin{equation}\label{eqeqeqeq}
\tilde{Y}^\pi_{t_{i-1}} = \E\left[\left. Y^\pi_{t_{n}} + \sum_{k=i}^n f(X^{\pi}_{t_{k-1}},\tilde{Y}^{\pi}_{t_{k-1}},Z^{\pi}_{t_{k-1}})\Delta_k\right| X^{\pi}_{t_{i-1}}  \right]
\end{equation}
for an implicit method.

For an alternative type of simulation schemes, recall that for Markovian forward-backward equations, both $Y_t$ and $Z_t$ may be written as functions of the current forward state $(t,X_t)$. Hence, we use the regression estimator of (\ref{PIeqn16}) to write
\begin{equation}\label{PIeqn27}
\begin{aligned}
Y^\pi_{t_{i-1}} &= \hat{E}\left[\left. Y^\pi_{t_i}+f(X^{\pi}_{t_{i-1}},Y^{\pi}_{t_{i}},Z^{\pi}_{t_{i-1}})\Delta_i\right| X^{\pi}_{t_{i-1}} \right]\\
 &\equiv \hat{y}_{i-1}(X^{\pi}_{t_{i-1}})
\end{aligned}
\end{equation}
that is, the function $y(t,x)$ that generates $Y_t=y(t,X_t)$ is approximated with $\hat{y}(\cdot)$. Further, if we use $Z^\pi_{t_i}$ in the driver of (\ref{PIeqn27}) ($Z^\pi_{t_i}$ from the previous time-step) to obtain $\hat{y}_{i-1}(\cdot)$, we get the following scheme
\begin{equation}\label{PIeqn30}
\begin{aligned}
&Y^\pi_{t_n} = g(X^\pi_{t_{n}}),\,\,\,\,\, Z^\pi_{t_n} = \partial_x g(X^\pi_{t_{n}}) \sigma(X^\pi_{t_n}),\\
&Y^\pi_{t_{i-1}} = \hat{E}\left[\left. Y^\pi_{t_i}+f(X^{\pi}_{t_{i-1}},Y^{\pi}_{t_{i}},Z^{\pi}_{t_{i}})\Delta_i\right| X^{\pi}_{t_{i-1}} \right] \implies  \hat{y}_{i-1}(\cdot),\\
&Z^\pi_{t_{i-1}} = \partial_x \hat{y}_{i-1}(X^{\pi}_{t_{i-1}} )\sigma(X^{\pi}_{t_{i-1}} ),
\end{aligned}
\end{equation}
since the function that generates $Z_t$ is given by $Z_t = \partial_x y(t,X_t)\sigma(X_t)$, see footnote \ref{PIfn1}. In particular, if we employ the MARS regression, $\hat{y}(\cdot)$  will be a sum of piecewise linear splines and products thereof up to the specified degree. Hence, the partial derivatives $(\partial_s\hat{y}(\cdot),\partial_v\hat{y}(\cdot))$ are easily calculated analytically. Further, the last two calculations of (\ref{PIeqn30}) may be iterated with $Y^\pi_{t_{i-1}},Z^\pi_{t_{i-1}}$ for an implicit version of the scheme.

For a second type of modifications, we may include additional predictors for the regression functions. As an example, let $\Ch(t,x)$ denote the pricing function of an option with terminal payoff $g(X_T)$ calculated under Heston's model. As the pricing bound $Y_t$ lies in a neighbourhood of the price $\Ch(t,X_t)$, we may add this as a predictor to our regression estimator
\begin{equation}\label{eqeqqeqeqq}
Y^\pi_{t_{i-1}} = \hat{E}\left[\left. Y^\pi_{t_i}+f(X^{\pi}_{t_{i-1}},Y^{\pi}_{t_{i}},Z^{\pi}_{t_{i-1}})\Delta_i\right| X^{\pi}_{t_{i-1}} , \Ch(t_{i-1},X^{\pi}_{t_{i-1}}) \right]. 
\end{equation}

Finally, we mention a modification of the first regression in the standard scheme (\ref{PIeqn14}), as proposed by \cite{alanko2013reducing}
\begin{equation}\label{PIeqn20}
Z^\pi_{t_{i-1}} = \frac{1}{\Delta_i} \E\left[\left. \left(Y^\pi_{t_i}-\E[Y^\pi_{t_i}|X^{\pi}_{t_{i-1}}]\right)\Delta W_{t_i}\right| X^{\pi}_{t_{i-1}}  \right]
\end{equation}
with the purpose being a variance reduction of the regression estimate. The motivation is the following: since $Y^{\pi}_{t_i}=y(t_i,X^{\pi}_{t_i})$ 
for some continuous function $y(t,x)$, we have $Z^\pi_{t_{i-1}} = 
\E[y(t_{i-1}+\Delta_i,X^{\pi}_{t_{i-1}}+\Delta X^{\pi}_{t_i})\Delta W_{t_i}/\Delta_i|X^{\pi}_{t_{i-1}}]$ and the estimator thereof
\begin{equation}\label{PIeqn21}
\frac{1}{N}\sum_{j=1}^{N} y(t_{i-1}+\Delta_i,X^{\pi}_{t_{i-1}}+\Delta X^{\pi(j)}_{t_i})\frac{\sqrt{\Delta_i}z^{(j)}}{{\Delta_i}}
\end{equation}
where $z^{(j)}$ are independent standard normal random variables. As $\Delta X^{\pi(j)}_{t_i}= \text{drift}\times \Delta_i + \text{diff}\times\sqrt{\Delta_i}z^{(j)}$, we have that the variance of the estimate
(\ref{PIeqn21}) is approximately $y(t_{i-1},X^{\pi}_{t_{i-1}})^2/(N\Delta_i)$ for small $\Delta_i$ and hence, it blows up as $\Delta_i\rightarrow0$. In return if we use
\begin{equation}\label{PIeqn22}
\frac{1}{N}\sum_{j=1}^{N} \left( y(t_{i},X^{\pi}_{t_{i}}) - y(t_{i-1},X^{\pi}_{t_{i-1}})+f_{i-1}\Delta_i \right)\frac{\sqrt{\Delta_i}z^{(j)}}{{\Delta_i}}
\end{equation}
where $f_{i-1}\equiv f(X^{\pi}_{t_{i-1}},Y^{\pi}_{t_{i-1}},Z^{\pi}_{t_{i-1}})$ and $y(t_{i-1},X^{\pi}_{t_{i-1}})-f_{i-1}\Delta_i = \E[Y^\pi_{t_i}|X^{\pi}_{t_{i-1}}]$ from (\ref{PIeqn14}), we have that the estimator (\ref{PIeqn22}) of (\ref{PIeqn20}) will have  approximate variance $2y_x(t_{i-1},X^{\pi}_{t_{i-1}})^2/N+\Delta_i f_{i-1}/N$ which do not depend on $\Delta_i$ as this goes to zero.

We end this section with a demonstration of the simulation schemes based on (\ref{PIeqn16}) and (\ref{PIeqn17}) in the following example.

\begin{example}\label{PIex1}
For the forward process, we simulate $N=100,000$ paths of Heston's model (\ref{PIeqn13}) with parameters $(r,\kappa,\theta,\sigma,\rho)=(0,5.07,0.0457,0.48,-0.767)$, initial value $(S^{\pi}_0,V^{\pi}_0)=(100,\theta)$ over an equidistant time grid with $n=25$ points and terminal time $T=1$. For the backward process, we consider the trivial driver $f(X_t,Y_t,Z_t)=0$, i.e. $dY_t = Z_td\tilde{W}_t$, together with the terminal condition $Y_T = S_T$. Hence, $Y$ is a martingale and we have
\[
Y_t = \E^Q\left[S_T |\mathcal{F}_t \right] = S_t
\]
since for a zero interest rate, $S$ is a $Q$-martingale as well. As there is no dependency of $Z$ in the driver, the backward induction simplifies to the regression $Y^\pi_{t_{i-1}} = \hat{E}[ Y^\pi_{t_i}| X^{\pi}_{t_{i-1}} ]$ repeated for $i=n,\dots,1$ and a starting value $Y^\pi_{t_{n}} = S^{\pi}_{t_n}$. With $k=5$ nearest neighbours of the regression estimator (\ref{PIeqn15}), the left pane of Figure \ref{PIfig2} shows five simulated paths of the backward process $Y^{\pi}$ with the explicit scheme (\ref{PIeqn16}) along with the corresponding paths of $S^{\pi}$ and it can be seen that the components follow each other quite closely. Looking at the initial time value, the $N$-sample of $Y^{\pi}_0$ has an average 98.532 to be compared with the true value $Y_0 = \E^Q\left[S_T \right]=S_0=100$,  while the sample of $Y^{\pi}_{t_n}=S^{\pi}_{t_n}$ averages to 99.998.

\begin{figure}[!t]
\centering
\includegraphics[scale=0.43,trim=0 0 20 20,clip]{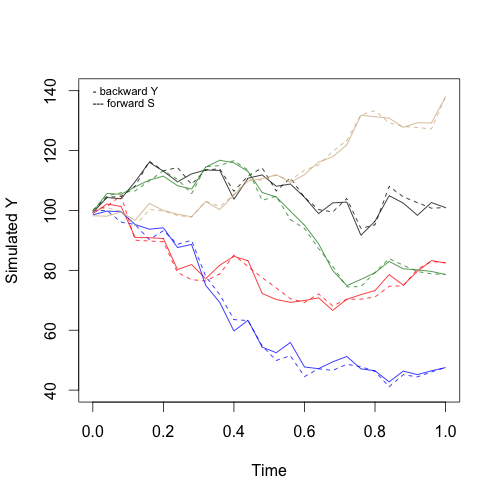}
\includegraphics[scale=0.43,trim=0 0 20 20,clip]{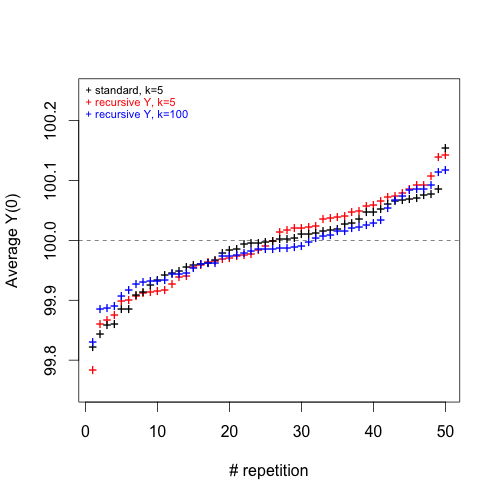}
\caption{\textbf{Left figure:} five simulated paths of $Y^{\pi}$  (solid lines) with the zero-driver of example \ref{PIex1}. The explicit scheme (\ref{PIeqn15})-(\ref{PIeqn16}) is employed with $k=5$ nearest neighbours. The forward component $X^{\pi}=(S^{\pi},V^{\pi})$ is simulated from Heston's model and the dashed lines show the corresponding paths of $S^{\pi}$. \textbf{Right figure:} the $N$-sample average of $Y^{\pi}_0$ (in increasing order) from 50 repetitions of the simulation with the $k=5$ explicit scheme (\ref{PIeqn16}) (black crosses), the recursive-based scheme (\ref{PIeqn17}) with $k=5$ (red crosses) and  $k=100$ (blue crosses).}
\label{PIfig2}
\end{figure}

If we repeat the simulation 50 times and calculate the average of $Y^{\pi}_0$ for each repetition, we obtain the result in the right pane of Figure \ref{PIfig2}. The first explicit scheme based on (\ref{PIeqn16}) yields sample averages quite close to the true value and if we repeat the simulations with the $Y^\pi$-recursion scheme (\ref{PIeqn17}) instead, we obtain similar results. For comparison, we have included the recursive scheme with $k=100$ nearest neighbours as well.

Finally, notice that this example corresponds to $g(x)=x$ and an effect $\alpha(S_t,V_t,u_t)=(0,0)^\top$ such that $Q^u\equiv Q$. Hence, with $\varrho=0$ for the driver (\ref{PIeqn10}), we have that the value process $J_t(u)= \E_u\left[g(S_T) |\mathcal{F}_t \right]= \E^Q\left[S_T |\mathcal{F}_t \right]$ is the solution to (\ref{PIeqn1}). 
\end{example}

\subsection{Simulation results for European options}\label{AppHe}
For numerical calculation of the pricing bounds for European options, we consider the parameter setting given in Table \ref{PItab5} and a set of call options with strike-maturity structure as given in Table \ref{PItab6}. The call prices are calculated from the so called semi-closed pricing formula of Heston's model, i.e. by numerical integration of the inverse Fourier transform of the price (see e.g. \cite{gatheral2011volatility}). The corresponding implied volatilities are then obtained from Black-Scholes formula by numerical optimisation. 

\begin {table}[h!]
\begin{center}
\begin{tabular}{| c |  c  c  c |} 
\multicolumn{4}{c}{\textbf{European call prices}} \\
\hline
\text{Strike/Expiry} & 75 & 100 & 125 \\
\hline
\hline
4m & 26.0044 (0.2823) & 4.8239 (0.2106) & 0.0070 (0.1518) \\
1y & 29.4915 (0.2482) & 10.9174 (0.2124) & 1.8403 (0.1832)\\
10y & 57.4959 (0.2220) & 46.4060 (0.2174) & 37.1943 (0.2138) \\
\hline
\end{tabular}
\end{center}
\caption{Prices and implied volatilities (in parenthesis) of European call options calculated by the semi-closed pricing formula of Heston's model with parameters from Table \ref{PItab5}.}
\label{PItab6}
\end{table}

Prior to considering the pricing bounds as obtained from the optimally controlled value process, we take a look at the prices one achieves by minimising/maximising Heston's pricing formula $C_{\text{He}}(\cdot)$ over the parameter uncertainty set $U$ represented by the elliptic constraint in (\ref{PIeqn99}) with a 95\% confidence level. That is
\begin{equation}\label{PIeqn25}
C_{\text{He}}^{\pm} = \left| \min_{(r,\kappa,\theta)\in U} \pm \Ch(S,V;\tau,K,\Theta) \right|
\end{equation}
where $\Theta$ is the vector of model parameters including $(r,\kappa,\theta)$ while $K$ is the strike and $\tau$ the time to maturity. From numerical optimisation of (\ref{PIeqn25}) with parameters and elliptic uncertainty region based on Table \ref{PItab5}, we get the results in Table \ref{PItab8}. We will use these  as a reference point for our forthcoming simulation study.

\begin {table}[h!]
\begin{center}
\begin{tabular}{| c |  c  c  c |} 
\multicolumn{4}{c}{\textbf{Optimised Heston pricing function}} \\
\hline
\text{Strike/Expiry} & 75 & 100 & 125 \\
\hline
\hline
 \multirow{ 2}{*}{4m}  	& $[25.9316,26.2591]$ & $[4.5758,5.0572]$ & $[0.0040,0.0124]$ \\
				& $(0.0520,0.3651)$  & $(0.1980,0.2225)$   & $(0.1441,0.1610)$\\
\hdashline[0.5pt/5pt]
 \multirow{ 2}{*}{1y} &    $[28.6578,30.4061]$ &  $[9.9716,11.8229]$  &  $[1.3840,2.4824]$\\
				& $(0.0303,0.3060)$  & $(0.1872,0.2364)$  & $(0.1659,0.2053)$ \\
\hdashline[0.5pt/5pt]
 \multirow{ 2}{*}{10y}  &  $[54.5102,62.3675]$  &  $[40.2004,51.9955]$  & $[30.7291,43.0811]$ \\
				& $(0.0195,0.3190)$  &  $(0.1085,0.2925)$  & $(0.1444,0.2754)$ \\
\hline
\end{tabular}
\end{center}
\caption{Prices and implied volatilities of European call options, calculated by numerical minimisation/maximisation of the Heston pricing formula over the parameters $(r,\kappa,\theta)$ constrained by the parameter uncertainty region.}
\label{PItab8}
\end{table}

\begin {table}[h!]
\begin{center}
\begin{tabular}{|c  c c  c  c  c  c|} 
\multicolumn{7}{c}{\textbf{Model parameters}} \\
\hline
$S_0$ & $V_0$ & $r$ & $\kappa$  & $\theta$ & $\sigma$ & $\rho$ \\
\hline
100 & 0.0457 & 0.05 & 5.070 & 0.0457 & 0.4800 & -0.767 \\ 
\hline
\end{tabular}\\
\vspace{5pt}
\begin{tabular}{| c | c  c  c|} 
\hline
 & $r$  & $\kappa$ & $\beta$ \\ 
 \hline
 $r$ & 2.5e-05 & 0 & 0 \\ 
 $\kappa$ & 0 &  0.25 & 0 \\ 
 $\beta$ & 0 & 0 & 1e-04 \\
\hline
\end{tabular}
\end{center}
\caption{Parameter setting and covariance matrix used for the numerical calculation of pricing bounds for European options.}
\label{PItab5}
\end{table}

\subsubsection*{Simulation of the forward component}
The forward component $X=(S,V)$ of the SDE (\ref{PIeqn13}) governing the asset price and variance is simulated in the first stage of the simulation scheme for the forward-backward equation. 
We employ the standard Euler-Maruyama scheme for the log-price and an implicit Milstein scheme to generate the variance
\begin{equation*}
\begin{aligned}
\log S^{\pi}_{t_i} &=  \log S^{\pi}_{t_{i-1}} + \left(\mu-\frac{1}{2} V^{\pi}_{t_{i-1}}\right)\Delta_i + \sqrt{ V^{\pi}_{t_{i-1}} }
(\rho\Delta W^{1}_{t_i} + \sqrt{1-\rho^2}\Delta W^{2}_{t_i}) \\
V^{\pi}_{t_i} &= \frac{V^{\pi}_{t_{i-1}} + \kappa\theta\Delta_i + \sigma\sqrt{V^{\pi}_{t_{i-1}}} \Delta W^2_{t_{i}} + \frac{1}{4}\sigma^2((\Delta W^2_{t_{i}})^2-\Delta_i)}{1+\tilde{\kappa}\Delta_i}
\end{aligned}
\end{equation*} 
where $\Delta W^1_{t_i}$, $\Delta W^{2}_{t_i}$ are independent variables generated from the zero-mean normal distribution with variance $\Delta_i$. If the parameters satisfy $4\kappa\theta>\sigma^2$ this discretization scheme generates positive variance paths and we do not have to impose any truncation as in the case with the standard Euler-Maruyama scheme, see \cite{andersen2010} or \cite{alfonsi2016affine}. We simulate  $N=100,000$ paths over an equidistant time gird with $n=25$ knots. 

\subsubsection*{The optimised Heston formula by backward simulation}
As a first simulation example of the backward component $Y$, we consider the formula-optimal price of the at-the-money call with maturity one year (prices given in Table \ref{PItab8}). Hence, we simulate the backward component with the non-optimised driver $f(X_t,Y_t,Z_t,u_t)$ of equation (\ref{PIeqn10}) with a constant $u_t$ based on the resulting parameters from the price-optimisation (\ref{PIeqn25}) of the considered call option. This allows us to evaluate the accuracy of our simulations schemes in a situation where we know the true values we are aiming at calculating numerically. The reason for simulating the optimised price of (\ref{PIeqn25}) instead of the null-controlled {plain} price of the call (given in Table \ref{PItab6}) is that the optimised-price simulation relies on a $Z$-dependent driver, while the $Q$-price has an effect being zero in (\ref{PIeqn10}) such that the $Z$-regression step of the simulation scheme expires for the plain price (cf. example \ref{PIex1}). 

\begin{figure}[!t]
\centering
\includegraphics[scale=0.43,trim=0 0 20 20,clip]{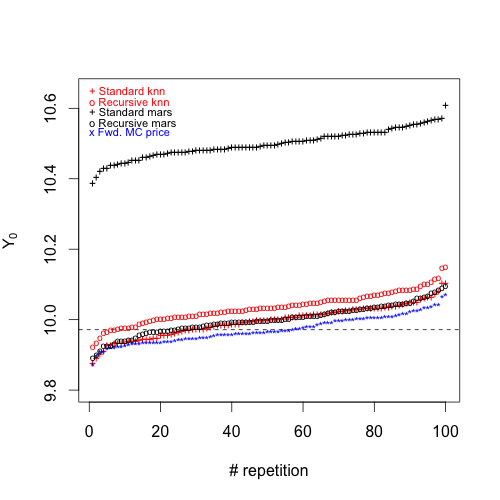}
\includegraphics[scale=0.43,trim=0 0 20 20,clip]{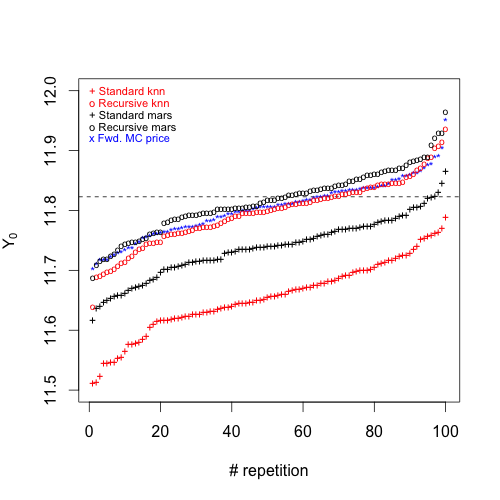}
\caption{Numerical calculation of the formula-optimal price of the one-year at-the-money call (Table \ref{PItab8}). \textbf{Left figure:} the $N$-sample average of the minimised price $Y^{\pi}_0$ from 100 repetitions of the simulation (in increasing order). We use a equidistant time-grid with $n=25$ time-points and generate $N=100,000$ paths of $Y^{\pi}$ in every simulation. \textbf{Right figure:} the corresponding maximised price. The figures show the results from four explicit  schemes based on the $k=5$ nearest neighbours estimator (red marks) and the MARS estimator of degree 2 (black marks). The dashed lines indicate the true call price as calculated by the (optimised) Heston's formula while the blue stars show the Monte-Carlo price as calculated from the $N$ simulated paths of $X^{\pi}=(S^{\pi},V^{\pi})$ for each repetition.}
\label{PIfig3}
\end{figure}

For starters, we consider the following four variations of the simulation schemes from the previous section: 
\begin{itemize}
\item[\textbf{1}.] the explicit scheme (\ref{PIeqn16}) with $k=5$ nearest neighbours regression (\ref{PIeqn15})
\item[\textbf{2}.] the explicit scheme with MARS regression (\ref{PIeqn23}) of degree 2
\item[\textbf{3}.] the explicit-recursive scheme (\ref{PIeqn17}) with $k=5$ nearest neighbours regression
\item[\textbf{4}.] the explicit-recursive scheme with MARS regression of degree 2.
\end{itemize}
For each of the schemes \textbf{1}--\textbf{4}, we repeatedly simulate the formula-optimal price 100 times and calculate sample- bias and root mean square errors. The results are given in Table \ref{PItab9}, while Figure \ref{PIfig3} shows the prices from all repetitions of the simulation.

\begin {table}[h!]
\begin{center}
\begin{tabular}{| c |  c c  c |} 
\multicolumn{4}{c}{\textbf{Backward simulated optimised Heston price}} \\
\hline
\text{Scheme} &  Ave. ${\E}(\hat{\pi})$ &Bias: ${\E}(\hat{\pi})-\pi$ & RMSE: $\sqrt{\E[(\hat{\pi}-\pi)^2]}$  \\
\hline
\hline
Explicit knn 		& 11.6552 & -0.1677 & 0.1783\\
Explicit MARS		& 11.7378 & -0.0851 & 0.0987\\
Recursive knn 		& 11.7968 & -0.0261 & 0.0608\\
Recursive MARS	& 11.8164 & -0.0065 & 0.0534\\
Forward MC 		& 11.8041 & -0.0188 & 0.0508\\
\hline\hline
Explicit knn 		& 9.9960   & 0.0244 & 0.0511\\
Explicit MARS		& 10.4993 & 0.5277 & 0.5292\\
Recursive knn 		& 10.0351 & 0.0635 & 0.0766\\
Recursive MARS	& 10.0004 & 0.0288 & 0.0509\\
Forward MC 		& 9.9719   & 0.0003 & 0.0380\\
\hline
\end{tabular}
\end{center}
\caption{Accuracy of the simulated formula-optimised price of an at-the-money call option with maturity one year (true values in Table \ref{PItab8}) for $N=100,000$ and $n=25$. Sample- average, bias and root mean square error calculated from 100 repetitions of each simulation.}
\label{PItab9}
\end{table}

From Table \ref{PItab9} we see that the explicit-recursive-MARS scheme performs best in terms of low bias and low RMSE although the simple explicit-knn scheme performs well for the lower price. Comparing the backward simulation with the Monte Carlo price calculated directly from forward simulation we have close to equal performance for the higher price. Since the backward simulation step is dependent of the forward step, we can not expect any improvement in accuracy beyond that of the forward simulation.

\begin {table}[h!]
\begin{center}
\begin{tabular}{| c |  c c  c |} 
\multicolumn{4}{c}{\textbf{Backward simulated optimised Heston price II}} \\
\hline
\text{Scheme} &   ${\E}(\hat{\pi})$ & ${\E}(\hat{\pi})-\pi$ & $\sqrt{\E[(\hat{\pi}-\pi)^2]}$  \\
\hline
\hline
Forward MC 		& 11.8094 & -0.0135 &  0.0411\\
\hdashline[0.5pt/5pt]
Rec. MARS, var. reduction & 11.8169 & -0.0060  & 0.0433  \\
Rec. MARS, two implicit & 11.8196 & -0.0033 &  0.0468 \\
Rec. MARS, var. red. \& two imp. & 11.8164 & -0.0065 & 0.0433 \\
Rec. MARS,  call-predictor  & 11.8166 & -0.0063 &  0.0435 \\
Rec. MARS, $Z$-function & 11.6868 & -0.1361 &  0.1489 \\
Rec. MARS, $Z$-fun. \& three imp. & 11.6794 & -0.1435 &  0.1558  \\
\hline\hline
Forward MC 		& 9.9719   & 0.0003 & 0.0380\\
\hdashline[0.5pt/5pt]
Rec. MARS, var. reduction & 10.0075 & 0.0359 & 0.0501 \\
Rec. MARS, two implicit & 10.0027 & 0.0311 & 0.0495 \\
Rec. MARS, var. red. \& two imp. & 10.0094 & 0.0378 & 0.0515 \\
Rec. MARS,  call-predictor  & 10.0082 & 0.0366 & 0.0507 \\
Rec. MARS, $Z$-function & 10.1096 & 0.1380 & 0.1502 \\
Rec. MARS, $Z$-fun. \& three imp. &  10.1661 & 0.1945 & 0.2034\\
\hline
\end{tabular}
\end{center}
\caption{Accuracy of the simulated formula-optimised price of an at-the-money call option with maturity one year (true values in Table \ref{PItab8}) for $N=100,000$ and $n=25$. Sample- average, bias and root mean square error calculated from 100 repetitions of each simulation.}
\label{PItab10}
\end{table}

Next, we continue with the following modifications of the simulation schemes:
\begin{itemize}
\item[\textbf{5}.] explicit-recursive-MARS with variance reduction (\ref{PIeqn20})
\item[\textbf{6}.] explicit-recursive-MARS with two implicit iterations (\ref{eqeqeqeq})
\item[\textbf{7}.] a combination of \textbf{5} and \textbf{6}
\item[\textbf{8}.] explicit-recursive-MARS with call-price predictor\footnote{The calculation of the pricing-formula for the call relies on numerical integration and we need $N=100,000$ such evaluations for each of $n=25$ time-step which makes the scheme very computer intensive. For this reason, we calculate a subset of 500 call prices and use a polynomial regression to predict the remaining call prices. As the pricing formula is a ``nice" function of $S$ and $V$, this approximation only has a limited impact.} (\ref{eqeqqeqeqq}).
\end{itemize}
The results are recorded in Table \ref{PItab10} and if we compare these with the result for the plain explicit-recursive-MARS scheme \textbf{4}, we observe similar accuracies for all of them.  However, as both implicit schemes \textbf{6} and \textbf{7} add $N$ regressions and $N$ evaluations of the driver to the computational cost for each implicit iteration, we opt for the schemes \textbf{4} or \textbf{5}. 

At last, we consider two schemes based on the MARS derivative: (\textbf{9}) explicit-recursive-MARS with $Z$-function (\ref{PIeqn30}), (\textbf{10}) explicit-recursive-MARS with $Z$-function and three implicit iterations.  Both these modifications yield poor accuracy, see Table \ref{PItab10}.

\subsection{The optimally controlled value-process of a European call option}\label{AppNS}
Here we simulate the backward component of equation (\ref{PIeqn2}) that governs the optimally controlled upper/lower pricing bound of the European call option with strike-maturity structure as in Table \ref{PItab6} based on parameters in Table \ref{PItab5}. Hence, we simulate $Y$ with an initial (terminal) condition $Y_{t_n}^{\pi}=(S^{\pi}_{t_n}-K)^+$ and an optimised driver $H(X_t,Y_t,Z_t)$ as of equation (\ref{PIeqn19}) with a confidence level of 95\% for the parameter uncertainty region based on the covariance matrix in Table \ref{PItab5}. 

As before, we simulate the forward component $X^{\pi}=(S^{\pi},V^{\pi})$ with the Euler-Maruyama implicit-Milstein scheme and for a start, we use $N=100,000$ paths over an equidistant time-gird with $n=25$ points. Note that for each backwards time-step, we perform $3\times N$ regressions to obtain the one-step recursion of $Z^{1\pi}$, $Z^{2\pi}$, $Y^{\pi}$ and $N$ evaluations of the matrix multiplication in (\ref{PIeqn19}) for the optimal driver. For the implicit versions of the schemes, we iterate two (or three) times which adds $2\times N$ regressions and $2\times N$ matrix multiplications to each time-step.

\begin{figure}[!t]
\centering
\includegraphics[scale=0.43,trim=0 0 20 20,clip]{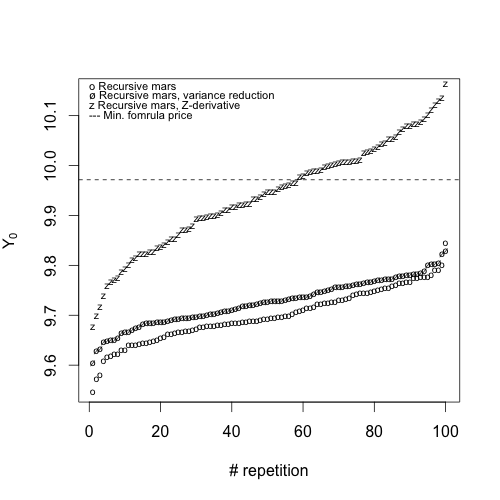}
\includegraphics[scale=0.43,trim=0 0 20 20,clip]{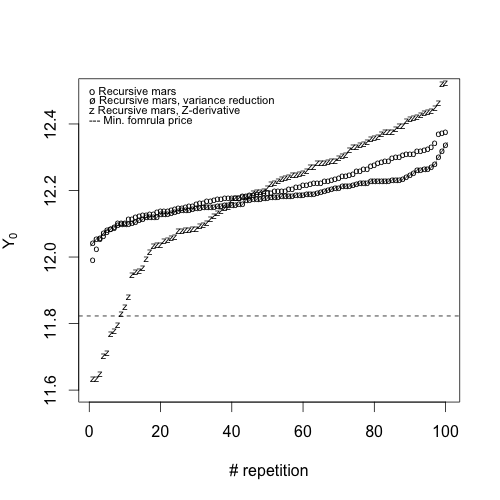}
\caption{Numerical calculation for the pricing bounds of the one-year at-the-money call with $N=100,000$ paths over $n=25$ time-points. \textbf{Left figure:} the $N$-sample average $Y^{\pi}_0$ of the lower bound for the call price (in increasing order) calculated from each of 100 repetitions of the simulation. \textbf{Right figure:} the corresponding upper bound. The dashed lines indicate the call price as calculated by the optimised Heston's formula.}
\label{PIfig4}
\end{figure}

For a demonstrative example, we again consider the one-year the at-the-money call option and run 100 repetitions of the following simulation schemes: 
\begin{itemize}
\item[\textbf{1}.] explicit-recursive-MARS of degree 2
\item[\textbf{2}.] explicit-recursive-MARS of degree 2 with variance reduction
\item[\textbf{3}.] explicit-recursive-MARS of degree 2 with $Z$ calculated from the MARS derivative
\item[\textbf{4}.] two implicit fixed-point iterations added to scheme number \textbf{1}
\item[\textbf{5}.] two implicit fixed-point iterations added to scheme number \textbf{2}
\item[\textbf{6}.] explicit-recursive-MARS of degree 2 with call-price predictor and variance reduction.
\end{itemize}
The resulting pricing bounds are shown in Figure \ref{PIfig4} where, for clarity, we have plotted only the results from scheme number \textbf{1}, \textbf{2} and \textbf{3}.

From Figure \ref{PIfig4} we see that if we add variance reduction to the explicit-recursive-MARS we obtain slightly higher (lower) prices for the lower (upper) boundary and a somewhat lower variance. Further, if we consider the two-step implicit versions of these schemes, we have that \textbf{1} and \textbf{4} coincide almost perfectly, and also \textbf{2} and \textbf{5}, for both the upper and lower bounds (these schemes are excluded from Figure \ref{PIfig4} only for clarity). The same holds if we add the call-price predictor: \textbf{2} and \textbf{6} coincide for both the upper and lower bounds. As in the case for the formula-optimised price, the $Z$-function scheme yields a high lower bound (similar to the formula minimised price) and a upper bound similar to the other schemes. Both bounds also have a very high variance, and for these reasons we henceforth omit the $Z$-function schemes.

\begin {table}[h!]
\begin{center}
\begin{tabular}{| c |  c  c  c |} 
\multicolumn{4}{c}{\textbf{Recursive MARS degree 2 with 
variance reduction, $n=25$}}\\
\hline
\text{Strike/Expiry} & 75 & 100 & 125 \\
\hline
\hline
 \multirow{ 2}{*}{4m}  & [25.7771, 26.2877] &  [4.5005,5.1597] &  [0.0016,0.0175] \\
				& (0.0585,0.3714) & (0.1942,0.2277) & (0.1335,0.1672) \\
\hdashline[0.5pt/5pt]
  \multirow{ 2}{*}{1y}  	& [28.5910,30.5482] & [9.7418,12.1603] &  [  1.2374,2.6306]   \\
				&  (0.0329,0.3138) & (0.1811,0.2454)  & (0.1600,0.2101)\\
\hdashline[0.5pt/5pt]
 \multirow{ 2}{*}{10y} &   [52.7314,64.9297] & [40.5299,54.2037]  & [30.7684,45.1219]   \\
				& (0.0319,0.3645) & ( 0.1176,0.3214)  & ( 0.1448,0.2970) \\
\hline
\end{tabular}
\end{center}
\caption{Pricing bounds for the European call option and corresponding Black-Scholes implied volatilities. Calculated from numerical simulation schemes for the backward process with $N=100,000$ simulated paths of the forward process following Heston's model over an equidistant time grid with $n=25$ points.}
\label{PItab11}
\end{table}

\begin{figure}[t!]
\centering
\includegraphics[scale=0.43,trim=0 20 20 20,clip]{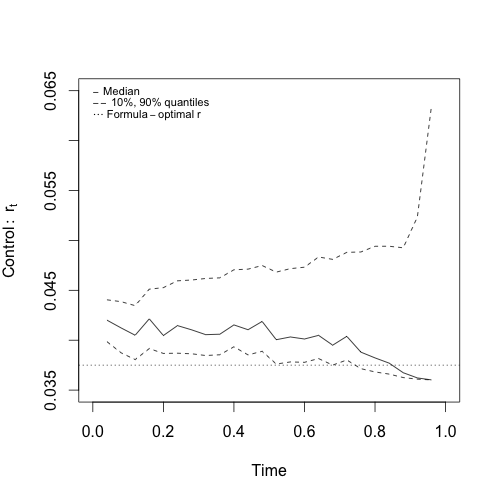}
\includegraphics[scale=0.43,trim=0 20 20 20,clip]{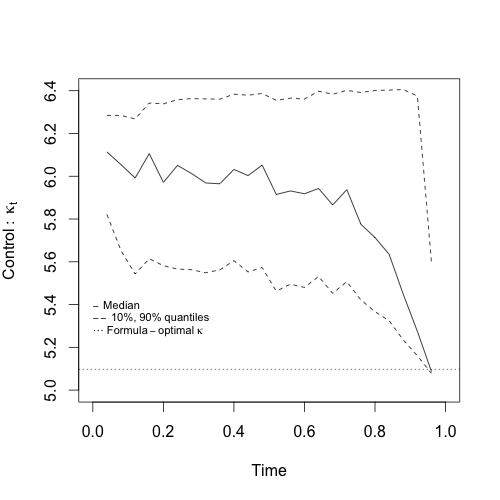}
\includegraphics[scale=0.43,trim=0 0 20 20,clip]{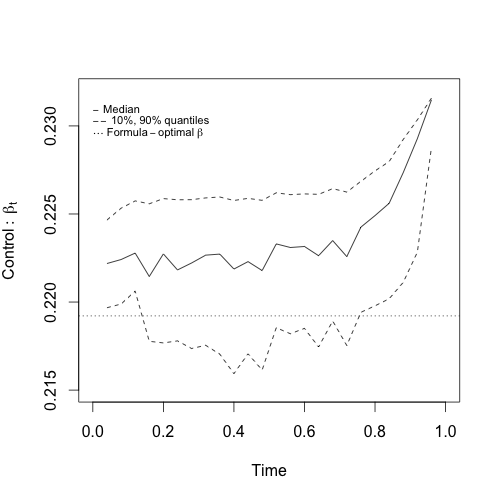}
\caption{The optimal controls $u^*_t=(r^*_t,\kappa^*_t,\beta^*_t)$ as outputted from the optimisation of the driver $H^-$ for the one-year ATM call option. Plotted median and quantiles of $N=100,000$ simulation paths. The dotted lines show the corresponding constant parameter choice from the optimised Heston formula.}
\label{PIfig5}
\end{figure}

Based on the previous results for the one-year ATM call, we choose to employ the explicit-recursive-MARS with variance reduction as our working scheme for pricing-boundary calculations. The simulation based results for the considered call options of Table \ref{PItab6} are given in Table \ref{PItab11} and if we compare these with the formula-optimal prices of Table \ref{PItab8}, we generally see wider pricing intervals for the optimally controlled value process. This is what we should expect: the formula-optimal prices correspond to a controlled value-process with parameters held constant throughout the lifetime of the option, while in the former case, the parameters are allowed to vary in an optimal way. An illustration of this point is given in Figure \ref{PIfig5} where it is shown how the parameters vary for the optimally controlled one-year at-the-money call option.

The previous pricing bounds where obtained from a simulation of $N=100,000$ paths over a regular time-grid of $n=25$ points. While $N$ is chosen to be a high number for the gain of a low error of the simulation-based regression estimator, the discretization time-step $\Delta=T/n$ is relatively large (for the one-year option, $n=25$ corresponds to having a time-step in the size of two weeks while for practical Monte Carlo pricing, one typically uses daily or even finer time-steps). For this reason we repeat the calculation of Table \ref{PItab11} with a finer time-step of $n=100$. The results given in Table \ref{PItab12} show wider pricing bounds for all strikes/maturities when comparing to tabel \ref{PItab11} and the difference between the two step sizes increases with the maturity. A natural explanation for this is that with a higher number of $n$ we also have a higher number of time-steps at which we optimise the driver $H^{\pm}$, and this should lead to a value process $Y^{\pi}$ optimised to a higher degree. This effect is obvious for the long-maturity options while it is less apparent for the four-month option (the implied volatilities agrees down to $10^{-2}$) which also indicates that the simulation error is not particularly affected by the finer time-discretization.

\begin {table}[h!]
\begin{center}
\begin{tabular}{| c |  c  c  c |} 
\multicolumn{4}{c}{\textbf{Recursive MARS degree 2 with 
variance reduction, $n=100$}}\\
\hline
\textit{Strike/Expiry} & 75 & 100 & 125 \\
\hline
\hline
 \multirow{ 2}{*}{4m}  & [25.7349, 26.3130] &  [4.4748, 5.1885] &  [0.0005054, 0.02127] \\
				& (0.05824, 0.3767) & (0.1929, 0.2292) & (0.1225, 0.1710) \\
\hdashline[0.5pt/5pt]
  \multirow{ 2}{*}{1y}  	& [28.3640, 30.6353] & [9.6158, 12.2530]  & [1.1033, 2.6726]   \\
				& (0.0330, 0.3184) & (0.1777, 0.2478) & (0.1543, 0.2115) \\
\hdashline[0.5pt/5pt]
 \multirow{ 2}{*}{10y} &  [48.5895, 67.3999] & [36.9068, 57.1310] &  [27.4504, 48.1860]\\
				&  (0.0217, 0.4076) &  (0.0199, 0.3598) &  (0.1053, 0.3298)\\
\hline
\end{tabular}
\end{center}
\caption{Pricing bounds for the European call option and corresponding Black-Scholes implied volatilities. Calculated from numerical simulation schemes for the backward process with $N=100,000$ simulated paths of the forward process following Heston's model over an equidistant time grid with $n=100$ points.}
\label{PItab12}
\end{table}

\end{appendix}

\clearpage
\bibliographystyle{apalike}
\bibliography{biblio}

\begin{thebibliography}{}

\bibitem[A\"it-Sahalia and Kimmel, 2007]{ai2007maximum}
A\"it-Sahalia, Y. and Kimmel, R. (2007).
\newblock Maximum likelihood estimation of stochastic volatility models.
\newblock {\em Journal of Financial Economics}, 83(2):413--452.

\bibitem[Alanko and Avellaneda, 2013]{alanko2013reducing}
Alanko, S. and Avellaneda, M. (2013).
\newblock Reducing variance in the numerical solution of \textsc{BSDE}s.
\newblock {\em Comptes Rendus Mathematique}, 351(3):135--138.

\bibitem[Alfonsi, 2016]{alfonsi2016affine}
Alfonsi, A. (2016).
\newblock {\em Affine Diffusions and Related Processes: Simulation, Theory and
  Applications}.
\newblock Springer.

\bibitem[Andersen et~al., 2010]{andersen2010}
Andersen, L.~B., Jäckel, P., and Kahl, C. (2010).
\newblock {\em Simulation of Square-Root Processes}.
\newblock John Wiley and Sons, Ltd.

\bibitem[Andersen and Benzoni, 2009]{andersen2009realized}
Andersen, T. and Benzoni, L. (2009).
\newblock Realized volatility.
\newblock {\em Handbook of Financial Time Series}, pages 555--575.

\bibitem[Avellaneda et~al., 1995]{avellaneda1995pricing}
Avellaneda, M., Levy∗, A., and Par{\'a}s, A. (1995).
\newblock Pricing and hedging derivative securities in markets with uncertain
  volatilities.
\newblock {\em Applied Mathematical Finance}, 2(2):73--88.

\bibitem[Avellaneda and Paras, 1996]{avellaneda1996managing}
Avellaneda, M. and Paras, A. (1996).
\newblock Managing the volatility risk of portfolios of derivative securities:
  the lagrangian uncertain volatility model.
\newblock {\em Applied Mathematical Finance}, 3(1):21--52.

\bibitem[Bann{\"o}r and Scherer, 2014]{bannor2014model}
Bann{\"o}r, K.~F. and Scherer, M. (2014).
\newblock Model risk and uncertainty - illustrated with examples from
  mathematical finance.
\newblock In {\em Risk - A Multidisciplinary Introduction}, pages 279--306.
  Springer.

\bibitem[Barczy and Pap, 2016]{barczy2016asymptotic}
Barczy, M. and Pap, G. (2016).
\newblock Asymptotic properties of maximum-likelihood estimators for {H}eston
  models based on continuous time observations.
\newblock {\em Statistics}, 50(2):389--417.

\bibitem[Bates, 1996]{bates1996jumps}
Bates, D.~S. (1996).
\newblock Jumps and stochastic volatility: Exchange rate processes implicit in
  deutsche mark options.
\newblock {\em Review of financial studies}, 9(1):69--107.

\bibitem[Bender and Denk, 2007]{bender2007forward}
Bender, C. and Denk, R. (2007).
\newblock A forward scheme for backward \textsc{SDE}s.
\newblock {\em Stochastic processes and their applications},
  117(12):1793--1812.

\bibitem[Bene{\v s}, 1970]{3}
Bene{\v s}, V. (1970).
\newblock Existence of optimal strategies based on specified information, for a
  class of stochastic decision problems.
\newblock {\em SIAM J. Control}, 8:179--188.

\bibitem[Bibby and S{\o}rensen, 1995]{bibby1995martingale}
Bibby, B.~M. and S{\o}rensen, M. (1995).
\newblock Martingale estimation functions for discretely observed diffusion
  processes.
\newblock {\em Bernoulli}, pages 17--39.

\bibitem[Black, 1976]{black1976stuedies}
Black, F. (1976).
\newblock Studies of stock price volatility changes.
\newblock {\em In: Proceedings of the 1976 Meetings of the American Statistical
  Association}, pages 171--181.

\bibitem[Black and Scholes, 1973]{black1973pricing}
Black, F. and Scholes, M. (1973).
\newblock The pricing of options and corporate liabilities.
\newblock {\em The journal of political economy}, pages 637--654.

\bibitem[Bouchard and Elie, 2008]{bouchard2008discrete}
Bouchard, B. and Elie, R. (2008).
\newblock Discrete-time approximation of decoupled forward--backward
  \textsc{SDE} with jumps.
\newblock {\em Stochastic Processes and their Applications}, 118(1):53--75.

\bibitem[Bouchard and Touzi, 2004]{bouchard2004discrete}
Bouchard, B. and Touzi, N. (2004).
\newblock Discrete-time approximation and monte-carlo simulation of backward
  stochastic differential equations.
\newblock {\em Stochastic Processes and their applications}, 111(2):175--206.

\bibitem[Breeden, 1979]{breeden1979intertemporal}
Breeden, D.~T. (1979).
\newblock An intertemporal asset pricing model with stochastic consumption and
  investment opportunities.
\newblock {\em Journal of Financial Economics}, 7(3):265--296.

\bibitem[Bunnin et~al., 2002]{bunnin2002option}
Bunnin, F.~O., Guo, Y., and Ren, Y. (2002).
\newblock Option pricing under model and parameter uncertainty using predictive
  densities.
\newblock {\em Statistics and Computing}, 12(1):37--44.

\bibitem[Carr et~al., 2003]{carr2003stochastic}
Carr, P., Geman, H., Madan, D.~B., and Yor, M. (2003).
\newblock Stochastic volatility for l{\'e}vy processes.
\newblock {\em Mathematical Finance}, 13(3):345--382.

\bibitem[Carr and Sun, 2007]{carr2007new}
Carr, P. and Sun, J. (2007).
\newblock A new approach for option pricing under stochastic volatility.
\newblock {\em Review of Derivatives Research}, 10(2):87--150.

\bibitem[Cohen and Elliott, 2015]{cohen2015stochastic}
Cohen, S.~N. and Elliott, R.~J. (2015).
\newblock {\em Stochastic Calculus and Applications}.
\newblock Springer.

\bibitem[Cont and Tankov, 2004]{contfinancial}
Cont, R. and Tankov, P. (2004).
\newblock {\em Financial modelling with jump processes}.
\newblock Chapman \& Hall/CRC.

\bibitem[Cox et~al., 1985]{cox1985theory}
Cox, J.~C., Ingersoll~Jr, J.~E., and Ross, S.~A. (1985).
\newblock A theory of the term structure of interest rates.
\newblock {\em Econometrica: Journal of the Econometric Society}, pages
  385--407.

\bibitem[de~Chaumaray, 2015]{de2015weighted}
de~Chaumaray, M. D.~R. (2015).
\newblock Weighted least squares estimation for the subcritical {H}eston
  process.
\newblock {\em arXiv preprint arXiv:1509.09167}.

\bibitem[Derman, 1996]{dermanMR}
Derman, E. (1996).
\newblock Model risk.
\newblock {\em Risk Magazine}, 9(5):34--37.

\bibitem[El~Karoui et~al., 1997]{El1997}
El~Karoui, N., Peng, S., and Quenez, M. (1997).
\newblock Backward stochastic differential equations in finance.
\newblock {\em Mathematical Finance}, 7(1):1--71.

\bibitem[El~Karoui and Quenez, 1995]{el1995dynamic}
El~Karoui, N. and Quenez, M.-C. (1995).
\newblock Dynamic programming and pricing of contingent claims in an incomplete
  market.
\newblock {\em SIAM journal on Control and Optimization}, 33(1):29--66.

\bibitem[Feller, 1951]{feller1951two}
Feller, W. (1951).
\newblock Two singular diffusion problems.
\newblock {\em Annals of mathematics}, pages 173--182.

\bibitem[Filippov, 1959]{Filippov}
Filippov, A. (1959).
\newblock On certain questions in the theory of optimal control.
\newblock {\em Vestnik Moskov. Univ. Ser. Mat. Meh. Astronom.}, 2:25--42.
\newblock English trans. J. Soc. Indust. Appl. Math. Ser. A. Control 1 (1962),
  76-84.

\bibitem[Gatheral, 2011]{gatheral2011volatility}
Gatheral, J. (2011).
\newblock {\em The volatility surface: a practitioner's guide}, volume 357.
\newblock John Wiley \& Sons.

\bibitem[Gobet, 2006]{Gobet06numericalsimulation}
Gobet, E. (2006).
\newblock Numerical simulation of \textsc{BSDE}s using empirical regression
  methods: theory and practice.
\newblock In {\em In Proceedings of the Fifth Colloquium on BSDEs (29th May -
  1st June 2005, Shangai) - Available on
  http://hal.archives-ouvertes.fr/hal00291199/fr}.

\bibitem[Gobet et~al., 2005]{Gobet2005}
Gobet, E., Lemor, J.-P., and Warin, X. (2005).
\newblock A regression-based monte carlo method to solve backward stochastic
  differential equations.
\newblock {\em The Annals of Applied Probability}.

\bibitem[Gobet and Turkedjiev, 2016]{gobet2016linear}
Gobet, E. and Turkedjiev, P. (2016).
\newblock Linear regression \textsc{MDP} scheme for discrete backward
  stochastic differential equations under general conditions.
\newblock {\em Mathematics of Computation}, 85(299):1359--1391.

\bibitem[Godambe and Heyde, 1987]{godambe1987quasi}
Godambe, V. and Heyde, C. (1987).
\newblock Quasi-likelihood and optimal estimation.
\newblock {\em International Statistical Review/Revue Internationale de
  Statistique}, pages 231--244.

\bibitem[Gupta et~al., 2010]{gupta2010model}
Gupta, A., Reisinger, C., and Whitley, A. (2010).
\newblock Model uncertainty and its impact on derivative pricing.
\newblock {\em Rethinking risk measurement and reporting}, 122.

\bibitem[Hastie et~al., 2005]{hastie2005elements}
Hastie, T., Tibshirani, R., Friedman, J., and Franklin, J. (2005).
\newblock The elements of statistical learning: data mining, inference and
  prediction.
\newblock {\em The Mathematical Intelligencer}, 27(2):83--85.

\bibitem[Heston, 1993]{heston1993closed}
Heston, S.~L. (1993).
\newblock A closed-form solution for options with stochastic volatility with
  applications to bond and currency options.
\newblock {\em Review of financial studies}, 6(2):327--343.

\bibitem[Heston, 1997]{heston1997simple}
Heston, S.~L. (1997).
\newblock A simple new formula for options with stochastic volatility.

\bibitem[Hull and White, 1987]{hull1987pricing}
Hull, J. and White, A. (1987).
\newblock The pricing of options on assets with stochastic volatilities.
\newblock {\em The journal of finance}, 42(2):281--300.

\bibitem[Jacquier and Jarrow, 2000]{jacquier2000bayesian}
Jacquier, E. and Jarrow, R. (2000).
\newblock Bayesian analysis of contingent claim model error.
\newblock {\em Journal of Econometrics}, 94(1):145--180.

\bibitem[Kessler, 1997]{kessler1997estimation}
Kessler, M. (1997).
\newblock Estimation of an ergodic diffusion from discrete observations.
\newblock {\em Scandinavian Journal of Statistics}, 24(2):211--229.

\bibitem[Keynes, 1921]{keynes1921treatise}
Keynes, J.~M. (1921).
\newblock A treatise on probability.

\bibitem[Kloeden and Platen, 1992]{kloeden1992numerical}
Kloeden, P.~E. and Platen, E. (1992).
\newblock {\em Numerical solution of stochastic differential equations},
  volume~23.
\newblock Springer.

\bibitem[Knight, 1921]{knight1921risk}
Knight, F.~H. (1921).
\newblock Risk, uncertainty and profit.
\newblock {\em Boston: Houghton Mifflin}.

\bibitem[Lehmann and Romano, 2006]{lehmann2006testing}
Lehmann, E.~L. and Romano, J.~P. (2006).
\newblock {\em Testing statistical hypotheses}.
\newblock Springer Science \& Business Media.

\bibitem[Lyons, 1995]{lyons1995uncertain}
Lyons, T.~J. (1995).
\newblock Uncertain volatility and the risk-free synthesis of derivatives.
\newblock {\em Applied mathematical finance}, 2(2):117--133.

\bibitem[Mc{S}hane and Warfield, 1967]{McShane}
Mc{S}hane, E. and Warfield, Jr., R. (1967).
\newblock On {F}ilippov's implicit functions lemma.
\newblock {\em Proceedings of the American Mathematical Society}, 18:41--47.

\bibitem[{Milborrow. Derived from mda:mars by T. Hastie and R. Tibshirani.},
  2011]{earthpackage}
{Milborrow. Derived from mda:mars by T. Hastie and R. Tibshirani.}, S. (2011).
\newblock {\em earth: Multivariate Adaptive Regression Splines}.
\newblock R package.

\bibitem[Peng, 1992]{Peng1992}
Peng, S. (1992).
\newblock A generalized dynamic programming principle and
  {H}amilton-{J}acobi-{B}ellman equation.
\newblock {\em Stochastics and Stochastics Reports}, 38:119--134.

\bibitem[Prakasa-Rao, 1983]{prakasa1983asymptotic}
Prakasa-Rao, B. (1983).
\newblock Asymptotic theory for non-linear least squares estimator for
  diffusion processes.
\newblock {\em Statistics: A Journal of Theoretical and Applied Statistics},
  14(2):195--209.

\bibitem[Quenez, 1997]{quenez1997stochastic}
Quenez, M.-C. (1997).
\newblock {\em Stochastic control and \textsc{BSDE}s}.
\newblock Addison Wesley Longman, Harlow.

\bibitem[Rogers, 2001]{rogers2001relaxed}
Rogers, L. C.~G. (2001).
\newblock The relaxed investor and parameter uncertainty.
\newblock {\em Finance and Stochastics}, 5(2):131--154.

\bibitem[Stein and Stein, 1991]{stein1991stock}
Stein, E.~M. and Stein, J.~C. (1991).
\newblock Stock price distributions with stochastic volatility: an analytic
  approach.
\newblock {\em Review of financial Studies}, 4(4):727--752.

\bibitem[Stein, 1989]{stein1989overreactions}
Stein, J. (1989).
\newblock Overreactions in the options market.
\newblock {\em The Journal of Finance}, 44(4):1011--1023.

\bibitem[Tegn\'er and Roberts, 2017]{mtGPLV}
Tegn\'er, M. and Roberts, S. (2017).
\newblock A bayesian take on option pricing with gaussian processes.
\newblock Working paper.

\bibitem[Wong and Heyde, 2006]{wong2006changes}
Wong, B. and Heyde, C. (2006).
\newblock On changes of measure in stochastic volatility models.
\newblock {\em International Journal of Stochastic Analysis}, 2006.

\end{thebibliography}

\end{document}